\documentstyle[12pt]{article}

\advance \topmargin by -\headheight
\advance \topmargin by -\headsep

\evensidemargin \oddsidemargin
\def\ie{{\em i.e.}}
\def\ie{\hbox{\it i.e.}}

\def\CC{{\mathchoice
{\rm C\mkern-8mu\vrule height1.45ex depth-.05ex
width.05em\mkern9mu\kern-.05em}
{\rm C\mkern-8mu\vrule height1.45ex depth-.05ex
width.05em\mkern9mu\kern-.05em}
{\rm C\mkern-8mu\vrule height1ex depth-.07ex
width.035em\mkern9mu\kern-.035em}
{\rm C\mkern-8mu\vrule height.65ex depth-.1ex
width.025em\mkern8mu\kern-.025em}}}

\def\RR{{\rm I\kern-1.6pt {\rm R}}}

\def\ZZ{{\rm Z}\kern-3.8pt {\rm Z} \kern2pt}
\def\IB{\relax{\rm I\kern-.18em B}}
\def\ID{\relax{\rm I\kern-.18em D}}
\def\II{\relax{\rm I\kern-.18em I}}
\def\IP{\relax{\rm I\kern-.18em P}}

\def\np{Nucl. Phys.}
\def\pl{Phys. Lett.}
\def\prl{Phys. Rev. Lett.}
\def\pr{Phys. Rev.}

\def\jhep{J. High Energy Phys.}

\newcommand{\beq}{\begin{equation}}
\newcommand{\eeq}{\end{equation}}
\newcommand{\rc}{\nonumber\\}
\newcommand{\bear}{\begin{eqnarray}}
\newcommand{\eear}{\end{eqnarray}}

\def\to{\rightarrow}

\def\to{\rightarrow}

\newfont{\namefont}{cmr10}
\newfont{\addfont}{cmti7 scaled 1440}
\newfont{\boldmathfont}{cmbx10}
\newfont{\headfontb}{cmbx10 scaled 1728}
\renewcommand{\theequation}{{\rm\thesection.\arabic{equation}}}
\begin{document}
\begin{titlepage}

\begin{center} \Large \bf Supersymmetric probes on the conifold

\end{center}

\vskip 0.3truein
\begin{center}
Daniel Are\'an
\footnote{arean@fpaxp1.usc.es},
David E.  Crooks
\footnote{dcrooks@fpaxp1.usc.es}
and
Alfonso V. Ramallo
\footnote{alfonso@fpaxp1.usc.es}

\vspace{0.3in}

Departamento de F\'\i sica de Part\'\i culas, Universidad de
Santiago de Compostela \\
E-15782 Santiago de Compostela, Spain
\vspace{0.3in}

\end{center}
\vskip 1
truein

\begin{center}
\bf ABSTRACT
\end{center}
We study the supersymmetric embeddings of different D-brane probes in the 
$AdS_5\times T^{1,1}$ geometry. The main tool employed is kappa symmetry and the
cases studied include D3-, D5- and D7-branes. We find a family of three-cycles of
the $T^{1,1}$ space over which a D3-brane can be wrapped supersymmetrically and
we determine the field  content of the corresponding gauge theory duals. 
Supersymmetric configurations of D5-branes wrapping a two-cycle and of spacetime
filling D7-branes are also found. The configurations in which the entire 
$T^{1,1}$ space is wrapped by a D5-brane (baryon vertex) and a D7-brane are also
studied. Some other embeddings which break supersymmetry but are nevertheless stable
are also determined.

\vskip4.6truecm
%\leftline{US-FT-3/03 }
\leftline{hep-th/0408210 \hfill August 2004}
\smallskip
\end{titlepage}
\setcounter{footnote}{0}

%%%%%%%%%%%%%%%%%%%%%%%%%%%%%%%%%%%%%%%%%%%%%%%%%%%%%%%%%%%%%%%%%%%%%%%%%%%%%%
%%%%%                       M A I N   T E X T
%%%%%%%%%%%%%%%%%%%%%%%%%%%%%%%%%%%%%%%%%%%%%%%%%%%%%%%%%%%%%%%%%%%%%%%%%%%%%%

%%%%%%%%%%%%%%%%%%%%%%%%%%%%%%%%%%%%%%%
%%%%%%%%%%
%%%%%%%%%%   Section 1: Introduction
%%%%%%%%%%
%%%%%%%%%%%%%%%%%%%%%%%%%%%%%%%%%%%%%%%

\setcounter{equation}{0}
\section{Introduction}
\medskip
The AdS/CFT correspondence \cite{jm} relates large $N$ gauge theory to string
theory. In the limit in which the `t Hooft coupling $g^2_{YM}\,N$ becomes infinite,
one can neglect the massive modes of the string and take type IIB supergravity on
the string  theory side of the correspondence \cite{MAGOO}. It is nevertheless very
interesting to understand how Yang-Mills theory encodes the full features of string
theory.

A possible way to uncover stringy effects in Yang-Mills theory consists of adding
D-branes on the supergravity side and trying to find out the corresponding field
theory dual. This approach was pioneered by Witten in ref. \cite{Wittenbaryon}.
Indeed, in \cite{Wittenbaryon} Witten considered a D3-brane wrapped over a
topologically non-trivial cycle of the $AdS_5\times {\bf \RR\IP}^5$ background and
showed that this configuration is dual to certain operators of dimension $N$ of the
$SO(N)$ gauge theory, namely the ``Pfaffians". Another  example along the same
lines is provided by the so-called giant gravitons, which are rotating branes
wrapped over a topologically trivial cycle \cite{GST}. These branes are not
topologically stable: they are stabilized dynamically by their angular momentum. The
corresponding field theory  duals have been found in ref. \cite{BBNS}.

In this paper we will study D-brane probes in the so-called Klebanov-Witten model
\cite{KW}. This model is obtained by placing a stack of $N$ D3-branes at the tip of
a conifold. The D3-branes warp the conifold metric and the resulting geometry
becomes $AdS_5\times T^{1,1}$, with $N$ units of Ramond-Ramond flux threading the 
$T^{1,1}$ space. The corresponding dual field theory is a  four-dimensional 
${\cal N}=1$ superconformal field theory with gauge group $SU(N)\times SU(N)$ 
coupled to four chiral superfields in the bifundamental representation.

The effect of adding  different D-branes to the Klebanov-Witten background has
already been studied in several places in the literature.  In ref. 
\cite{GK} it was proposed that D3-branes wrapped over three-cycles of $T^{1,1}$ are
dual to dibaryon operators built out of products of $N$ chiral superfields (see also
refs. \cite{BHK}-\cite{HMcK} for more results on dibaryons in this model and in some
orbifold theories). Moreover, it was also shown in ref. \cite{GK} that a D5-brane
wrapped over a two-cycle of $T^{1,1}$ behaves as a domain wall in $AdS_5$. On the
other hand, as first proposed in ref. \cite{KK}, one can use D7-branes to add
dynamical flavor to the Klebanov-Witten model (see also refs.
\cite{KKW}-\cite{Sonnen}). A list of the stable D-branes in this background,
obtained with methods quite different from those employed here, has appeared in
ref. 
\cite{MS}. 

The main technique that we will employ to determine the supersymmetric embeddings of
the different D-brane probes in the $AdS_5\times T^{1,1}$ background is kappa
symmetry \cite{swedes}. This approach is based on the fact that there exists a
matrix $\Gamma_{\kappa}$ such that, if $\epsilon$ is a Killing spinor of the
background, only those embeddings for which $\Gamma_{\kappa}\epsilon=\epsilon$
preserve some supersymmetry of the background \cite{bbs}. The matrix 
$\Gamma_{\kappa}$ depends on the metric induced on the worldvolume of the probe
and, therefore, if the Killing spinors are known, the kappa symmetry condition
gives rise to a set of first-order differential equations whose solutions (if they 
exist) determine the supersymmetric embedding of the brane probe. For these
configurations the kappa symmetry equation introduces some additional conditions on
$\epsilon$, which are only satisfied by some subset of the Killing spinors. Thus,
the probe only preserves some fraction of the original supersymmetry of the
background. For all the solutions we will find here we will be able to identify  the
supersymmetry that they preserve. Moreover, we will verify that the corresponding
embeddings satisfy the equations of motion derived from the Dirac-Born-Infeld
action of the probe. Actually, in all the cases studied, we will establish a series
of BPS bounds for the energy, along the lines of those studied in ref. \cite{GGT},
which are saturated by the kappa symmetric embeddings.

Clearly, to carry out the program sketched above we need to have a detailed
knowledge of the Killing spinors of the  $AdS_5\times T^{1,1}$ background. In
particular, it would be very useful to find a basis of frame one-forms in which the
spinors do not depend on the coordinates of the $T^{1,1}$ space. It turns out that
this frame is provided very naturally when the conifold is obtained
\cite{gaugedsugra} from an uplifting of eight-dimensional gauged supergravity
\cite{ss}. In this frame the Killing spinors are characterized by simple algebraic
conditions, and one can systematically explore the solutions of the kappa symmetry
equation. 

The first case we will study is that of the supersymmetric embeddings of D3-brane
probes. We will find a general family of three-cycles which contains, as a
particular case, the one used in ref. \cite{GK} to describe the dual of the
dibaryonic operator. We will be able to identify the field theory content of the
operators dual to our D3-brane embeddings. Moreover, we will also find two-cycles
on which the D3-brane can be wrapped in such a way that the equations of motion are
satisfied and, despite of the fact that supersymmetry is completely broken, the
system is stable.

We will consider next D5-brane probes, for which we will be able to identify the
two-cycle on which the D5-brane must be wrapped to realize the domain wall of the
four-dimensional gauge theory. We will also verify that if we wrap the D5-brane
over the same three-cycles which made  the D3-brane supersymmetric, one gets a
non-supersymmetric  stable solution of the equations of motion of the D5-brane
probe. The baryon vertex for the Klebanov-Witten model, a D5-brane wrapped over the
entire $T^{1,1}$, will be also analyzed. We will argue that this configuration
cannot be supersymmetric. 

Our final case is that corresponding to D7-brane probes. We will first study
the spacetime filling configurations. In this case
we will be able to find a two-parameter family of supersymmetric
embeddings which, in particular, include those proposed in refs. \cite{KK,Ouyang} as
suitable to add flavor to this background. Our results confirm that these
configurations are kappa symmetric. We will also show that the D7-brane can wrap
the entire $T^{1,1}$ and preserve some supersymmetry.

This paper is organized as follows. In section 2 we review the basic features of
the Klebanov-Witten model. In particular, we give the explicit form of the Killing
spinors in the frame which is more adequate to our purposes. We also introduce in
this section the general form of the kappa symmetry matrix $\Gamma_{\kappa}$ and
discuss the general strategy to solve the $\Gamma_{\kappa}\epsilon=\epsilon$
equation. 

Section 3 is devoted to the analysis of the supersymmetric D3-brane embeddings.
After choosing a set of convenient worldvolume coordinates and an ansatz for the
scalar fields that determine the embedding, we will be able to find a pair of
first-order differential equations whose solutions determine the supersymmetric
wrappings of the D3-brane over a three-cycle. This pair of equations can be solved
in general after a change of variables which converts them into the Cauchy-Riemann
equations. Similar analysis are carried out for the D5-brane and D7-brane probes in
sections 4 and 5 respectively. Some other possible embeddings for D3-, D5- and
D7-branes are discussed in appendix A. In section 6 we summarize our results and
draw some conclusions. We have also included in appendix B the calculation of the
Killing spinors of the so-called Klebanov-Strassler background \cite{KS}, which
could serve as a starting point to generalize our results to backgrounds dual to
theories without conformal invariance.

\setcounter{equation}{0}
\section{The Klebanov-Witten model}
\medskip
The conifold is a non-compact Calabi-Yau threefold with a conical singularity. Its
metric can be written as $ds^2_{6}=dr^2\,+\,r^2\,ds^2_{T^{1,1}}$, where 
$ds^2_{T^{1,1}}$ is the metric of the $T^{1,1}$ coset 
$(SU(2)\times SU(2))/U(1)$, which is the base of the cone. The $T^{1,1}$ space is
an Einstein manifold whose metric can be written \cite{Candelas}
explicitly by using the fact that
it is an $U(1)$ bundle over $S^2\times S^2$. Actually, if $(\theta_1,\phi_1)$ and 
$(\theta_2,\phi_2)$ are the standard coordinates of the $S^2$'s and if
$\psi\in [0,4\pi)$ parametrizes the $U(1)$ fiber, the metric may be written as:
\beq
ds^2_{T^{1,1}}\,=\,{1\over 6}\,\sum_{i=1}^{2}\,
\big(\,d\theta_i^2\,+\,\sin^2\theta_i\,d\phi_i^2\,)\,+\,
{1\over 9}\,\big(\,d\psi\,+\,\sum_{i=1}^{2}\cos\theta_id\phi_i\,\big)^2\,\,.
\label{t11metric}
\eeq
The conifold can also be described as the locus of points  
in $\CC^4$ which satisfy the equation:
\beq
z_1\,z_2\,-\,z_3\,z_4\,=\,0\,\,,
\label{conifold}
\eeq
which obviously has an isolated conical singularity at the origin of $\CC^4$. 
The relation between the holomorphic coordinates $z_i$ with the angles and
$r$ is:
\bear
&&z_1=r^{3/2}\,e^{{i\over 2}\,(\psi-\phi_1-\phi_2)}\,
\sin{\theta_1\over 2}\,\sin{\theta_2\over 2}\,\,,
\,\,\,\,\,\,\,\,\,\,\,\,\,\,\,\,\,
z_2=r^{3/2}\,e^{{i\over 2}\,(\psi+\phi_1+\phi_2)}\,
\cos{\theta_1\over 2}\,\cos{\theta_2\over 2}\,\,,
\rc\rc
&&z_3=r^{3/2}\,e^{{i\over 2}\,(\psi+\phi_1-\phi_2)}\,
\cos{\theta_1\over 2}\,\sin{\theta_2\over 2}\,\,,
\,\,\,\,\,\,\,\,\,\,\,\,\,\,\,\,\,
z_4=r^{3/2}\,e^{{i\over 2}\,(\psi-\phi_1+\phi_2)}\,
\sin{\theta_1\over 2}\,\cos{\theta_2\over 2}\,\,.\rc
\label{holomorphic}
\eear
It is also interesting to find some combinations of the $z_i$'s which only depend
on the coordinates $(\theta_1,\phi_1)$ or $(\theta_2,\phi_2)$. Actually, from the
parametrization (\ref{holomorphic}) it is straightforward to prove that:
\beq
{z_1\over z_3}\,=\,{z_4\over z_2}\,=\,
e^{-i\phi_1}\,\tan{\theta_1\over 2}\,\,,
\,\,\,\,\,\,\,\,\,\,\,\,\,\,\,\,\,\,\,\,\,\,\,\,
{z_1\over z_4}\,=\,{z_3\over z_2}\,=\,
e^{-i\phi_2}\,\tan{\theta_2\over 2}\,\,.
\label{zratio}
\eeq

By adding four Minkowski coordinates to the conifold we obtain a Ricci flat
ten-dimensional metric. Let us now place in this geometry a stack of $N$ coincident
D3-branes extended along the Minkowski coordinates and located at the singular
point of the conifold. The resulting model is the so-called Klebanov-Witten (KW)
model. The corresponding near-horizon metric and Ramond-Ramond selfdual five-form
are given by:
\bear
ds^2_{10}&=&[h(r)]^{-{1\over 2}}\,dx^2_{1,3}\,+\,[h(r)]^{{1\over 2}}\,
\big(\,dr^2\,+\,r^2\,ds^2_{T^{1,1}}\,\big)\,\,,\rc\rc
h(r)&=&{L^4\over r^4}\,\,,\rc\rc
g_s\,F^{(5)}&=&d^4x\,\wedge dh^{-1}\,+\,{\rm Hodge\,\,\, dual}\,\,,\rc\rc
L^4&=&{27\over 4}\,\pi g_s N\alpha'^2\,\,.
\label{KW}
\eear
The gauge theory dual to the supergravity background (\ref{KW}) is an ${\cal N}=1$
superconformal field theory with some matter multiplets.  Actually, the metric in
(\ref{KW}) can be written as:
\beq
ds^2_{10}\,=\,{r^2\over L^2}\,dx^2_{1,3}\,+\,{L^2\over r^2}\,dr^2\,+\,
L^2\,ds^2_{T^{1,1}}\,\,,
\label{adspoincare}
\eeq
which corresponds to the $AdS_5\times T^{1,1}$ space. In order to exhibit the field
theory dual to this background, let us solve the conifold equation (\ref{conifold})
by introducing four homogeneous coordinates $A_1$, $A_2$, $B_1$ and $B_2$ as
follows:
\beq
z_1=A_1\,B_1\,\,,
\,\,\,\,\,\,\,\,\,\,\,\,\,\,
z_2=A_2\,B_2\,\,,
\,\,\,\,\,\,\,\,\,\,\,\,\,\,
z_3=A_1\,B_2\,\,,
\,\,\,\,\,\,\,\,\,\,\,\,\,\,
z_4=A_2\,B_1\,\,.
\label{homogeneous}
\eeq 
Following the analysis of ref. \cite{KW}, one can show that the dual superconformal
theory can be described as an ${\cal N}=1$ $SU(N)\times SU(N)$ gauge theory which
includes four ${\cal N}=1$ chiral multiplets, which can be identified with (the
matrix generalization of) the  homogeneous coordinates $A_1$, $A_2$, $B_1$ and
$B_2$.  The fields $A_1$ and $A_2$ transform in the $({\bf N}, {\bf \bar N})$
representation of the gauge group, while $B_1$ and $B_2$ transform in the 
$({\bf \bar  N}, {\bf  N})$ representation. These fields are coupled through an
exactly marginal superpotential $W$ of the form:
\beq
W\,=\,\lambda \,\epsilon^{ij}\,\,\epsilon^{kl}\,
tr(A_iB_kA_jB_l)\,\,,
\eeq
where $\lambda$ is a constant.
The R-charge of the $A$ and $B$ fields is $1/2$, whereas their conformal dimension 
is $3/4$.

\subsection{Killing spinors}

As argued in ref. \cite{KW}, the KW  model preserves eight supersymmetries (see
also ref. \cite{Kehagias}). Notice that this is in agreement with the ${\cal N}=1$
superconformal character of the corresponding dual field theory, which has four
ordinary supersymmetries and four superconformal ones.

To obtain the explicit form of the Killing spinors, one has to look at the
supersymmetry variations of the dilatino and gravitino (see eq. (\ref{sugra})). It
turns out that the final result of the calculation is greatly simplified if some
particular basis of the frame one-forms for the $T^{1,1}$ part of the metric is
chosen. In order to specify this basis, let us define three one-forms associated to
a two-sphere
\beq
\sigma^1=d\theta_1\,\,,
\,\,\,\,\,\,\,\,\,\,\,\,
\sigma^2=\sin\theta_1\,d\phi_1\,\,,
\,\,\,\,\,\,\,\,\,\,\,\,
\sigma^3=\cos\theta_1d\phi_1\,\,,
\label{sigmaoneforms}
\eeq
and three one-forms associated to a three-sphere:
\bear
w^1&=&\sin\psi\sin\theta_2 d\phi_2\,+\,\cos\psi d\theta_2\,\,,\rc
w^2&=&-\cos\psi\sin\theta_2 d\phi_2\,+\,\sin\psi d\theta_2\,\,,\rc
w^3&=&d\psi\,+\,\cos\theta_2 d\phi_2\,\,.
\label{woneforms}
\eear
After an straightforward calculation one can verify that
these forms satisfy
\beq
d\sigma^i=-{1\over 2}\epsilon_{ijk}\,\sigma^j\wedge\sigma^k\,\,,
\,\,\,\,\,\,\,\,\,\,\,\,
dw^i={1\over 2}\epsilon_{ijk}\,w^j\wedge w^k\,\,.
\eeq
Moreover, the $T^{1,1}$ metric  (\ref{t11metric}) can be rewritten as
\beq
ds^2_{T^{1,1}}\,=\,{1\over 6}\,\big(\,(\sigma^1)^2\,+\,
(\sigma^2)^2\,+\,(w^1)^2\,+\,(w^2)^2\,\big)\,+\,
{1\over 9}\,\big(\,w^3+\sigma^3\,\big)^2\,\,.
\eeq
This form of writing the $T^{1,1}$ metric is the one that arises naturally when
the conifold geometry is obtained \cite{gaugedsugra} in the framework of the
eight-dimensional gauged supergravity obtained from a Scherk-Schwarz 
reduction of eleven dimensional supergravity on a SU(2) group manifold \cite{ss}. In
this approach one starts with a domain wall problem in eight dimensions and looks
for BPS solutions of the equations of motion. These solutions are subsequently
uplifted to eleven dimensions, where they represent gravity duals of branes wrapping
non-trivial cycles. The topological twist needed to realize supersymmetry with
wrapped branes is implemented in this approach in a very natural way. As shown in
ref. \cite{gaugedsugra}, the conifold metric is obtained as the gravity dual of
D6-branes wrapping a holomorphic $S^2$ inside a K3 manifold. Moreover, from the
consistency of the reduction, the Killing spinors should not depend on the
coordinates of the group manifold and, actually, in the one-form basis we will use
they do not depend on any angular coordinate of the $T^{1,1}$ space. Accordingly,
let us consider the following frame for the ten-dimensional metric (\ref{KW}):
\bear
&&e^{x^{\alpha}}\,=\,{r\over L}\,\,dx^{\alpha}
\,\,,\,\,\,\,\,\,\,\, (\alpha=0,1,2,3)\,\,,
\,\,\,\,\,\,\,\,\,\,\,\,\,\,\,\,
e^{r}\,=\,{L\over r}\,\,dr\,\,,\rc\rc
&&e^{i}\,=\,{L\over \sqrt{6}}\,\,\sigma^i\,\,,
\,\,\,\,\,\,\,\, (i=1,2)\,\,,\rc\rc
&&e^{\hat i}\,=\,\,{L\over \sqrt{6}}\,\,w^i\,\,,
\,\,\,\,\,\,\,\, (i=1,2)\,\,,\rc\rc
&&e^{\hat 3}\,=\,{L\over 3}\,\,(\,w^3+\sigma^3\,)\,\,.
\label{frame}
\eear
Let us also define the matrix $\Gamma_{*}$ as:
\beq
\Gamma_{*}\equiv i\Gamma_{x^0x^1x^2x^3}\,\,.
\label{gamma*}
\eeq
Then, the Killing spinors for the type IIB background (\ref{KW}) take the following
form:
\beq
\epsilon\,=\,r^{{\Gamma_{*}\over 2}}\,\,
\Big(\,1\,+\,{\Gamma_r\over 2L^2}\,\,x^{\alpha}\,\Gamma_{x^{\alpha}}\,\,
(1\,-\,\Gamma_{*}\,)\,\Big)\,\,\eta\,\,,
\label{adsspinor}
\eeq
where $\eta$ is a constant  spinor satisfying
\beq
\Gamma_{12}\,\eta\,=\,i\eta\,\,,
\,\,\,\,\,\,\,\,\,\,\,\,\,\,\,\,\,\,
\Gamma_{\hat 1\hat 2}\,\eta\,=\,-i\eta\,\,.
\label{tspinor}
\eeq
In eq. (\ref{adsspinor}) we are parametrizing the dependence of $\epsilon$ on the
$AdS_5$ coordinates as in ref. \cite{LPT}. Notice that, as the matrix multiplying
$\eta$ in eq. (\ref{adsspinor}) commutes with  $\Gamma_{12}$ and $\Gamma_{\hat
1\hat 2}$,  the spinor $\epsilon$  also satisfies the  conditions (\ref{tspinor}),
namely:
\beq
\Gamma_{12}\,\epsilon\,=\,i\epsilon\,\,,
\,\,\,\,\,\,\,\,\,\,\,\,\,\,\,\,\,\,
\Gamma_{\hat 1\hat 2}\,\epsilon\,=\,-i\epsilon\,\,.
\label{tsprojections}
\eeq
It is clear from eqs. (\ref{adsspinor}) and (\ref{tspinor}) that our system is $1/4$
supersymmetric, \ie\ it preserves 8 supersymmetries, as it corresponds to
the supergravity dual of a ${\cal N}=1$ superconformal field theory in
four dimensions. Moreover, let us decompose the constant spinor $\eta$ according to
the different eigenvalues of the matrix $\Gamma_{*}$:
\beq
\Gamma_{*}\,\eta_{\pm}\,=\,\pm\eta_{\pm}\,\,.
\label{etamasmenos}
\eeq
Using this decomposition in eq. (\ref{adsspinor}) we obtain two types
of Killing spinors
\bear
\epsilon_{+}&=&r^{1/2}\,\eta_+\,\,,\rc\rc
\epsilon_{-}&=&r^{-1/2}\,\eta_-\,+\,{r^{1/2}\over L^2}\,\,
\Gamma_r\,x^{\alpha}\Gamma_{x^{\alpha}}\,\eta_-\,\,.
\label{chiraladsspinor}
\eear
Notice that the four spinors $\epsilon_{+}$ are independent of the
coordinates $x^{\alpha}$ and 
$\Gamma_{*}\,\epsilon_{+}\,=\,\epsilon_{+}$. On the contrary, the 
$\epsilon_{-}$'s do depend on the $x^{\alpha}$'s and are not eigenvectors
of $\Gamma_{*}$. The latter correspond to the four superconformal supersymmetries,
while the $\epsilon_{+}$'s are the ones corresponding to the ordinary ones.

It is also interesting to write the form of the Killing spinors when
global coordinates are used for the $AdS_5$ part of the metric. In these
coordinates the ten-dimensional metric takes the form:
\beq
ds^2_{10}\,=\,L^2\,\Big[\,-\cosh^2\rho \,\,dt^2\,+\,d\rho^2\,+\,
\sinh^2\rho\,\,d\Omega_3^2\,\Big]\,+\, L^2\,ds^2_{T^{1,1}}\,\,,
\label{globalads}
\eeq
where $d\Omega_3^2$ is the metric of a unit three-sphere parametrized by
three angles $(\alpha^1, \alpha^2,\alpha^3)$:
\beq
d\Omega_3^2\,=\,(d\alpha^1)^2\,+\,\sin^2\alpha^1\Big(\,
(d\alpha^2)^2\,+\,\sin^2\alpha^2\,(d\alpha^3)^2\,\Big)\,\,,
\eeq
with $0\le\alpha^1,\alpha^2\le \pi$ and $0\le\alpha^3\le 2\pi$. 
In order to write down the Killing spinors in these coordinates, let us
choose the following frame for the $AdS_5$ part of the metric:
\bear
&&e^{t}\,=\,L\cosh\rho\,dt\,\,,\,\,\,\,\,\,\,\,\,\,\,\,\,\,\,\,
e^{\rho}\,=\,Ld\rho\,\,,\rc\rc
&&e^{\alpha^1}\,=\,L\sinh\rho\,d\alpha^1\,\,,\rc\rc
&&e^{\alpha^2}\,=\,L\sinh\rho\,\sin\alpha^1\,d\alpha^2\,\,,\rc\rc
&&e^{\alpha^3}\,=\,L\sinh\rho\,\sin\alpha^1\,\sin\alpha^2\,d\alpha^3\,\,.
\eear
We will continue to use the same frame forms as in eq. (\ref{frame}) for
the 
$T^{1,1}$ part of the metric. If we now  define the matrix
\beq
\gamma_*\,\equiv\,\Gamma_{t}\,\Gamma_{\rho}\,
\Gamma_{\alpha^1\,\alpha^2\,\alpha^3}\,\,,
\eeq
then, the Killing spinors in these coordinates can be written as \cite{globalads}
\beq
\epsilon\,=\,
e^{-i\,{\rho\over 2}\,\Gamma_{\rho}\gamma_*}\,
e^{-i\,{t\over 2}\,\Gamma_{t}\gamma_*}\,
e^{-{\alpha^1\over 2}\,\Gamma_{\alpha^1 \rho}}\,
e^{-{\alpha^2\over 2}\,\Gamma_{\alpha^2 \alpha^1}}\,
e^{-{\alpha^3\over 2}\,\Gamma_{\alpha^3 \alpha^2}}\,\eta\,\,,
\label{globalspinor}
\eeq
where $\eta$ is a constant spinor which satisfies the same conditions as
in eq. (\ref{tspinor}). 

\subsection{Supersymmetric probes}

Let us consider a Dp-brane probe in the KW background (\ref{KW}) and let $\xi^{\mu}$
($\mu=0,\cdots ,p$) be a set of worldvolume  coordinates. If $X^{M}$ denote
ten-dimensional coordinates,  the Dp-brane embedding will be characterized by a set
of functions 
$X^{M}(\xi^{\mu})$. The induced metric on the worldvolume is
\beq
g_{\mu\nu}\,=\,\partial_{\mu} X^{M}\,\partial_{\nu} X^{N}\,G_{MN}\,\,,
\label{inducedmetric}
\eeq
where $G_{MN}$ is the ten-dimensional metric. Let us denote by
$E_{N}^{\underline{M}}$ the coefficients that appear in the expression of  the
frame one-forms $e^{\underline{M}}$ of the ten-dimensional metric  in terms of
the differentials of the coordinates, namely:
\beq
e^{\underline{M}}\,=\,E_{N}^{\underline{M}}\,dX^N\,\,.
\eeq
Then, the induced Dirac matrices on the worldvolume are defined as
\beq
\gamma_{\mu}\,=\,\partial_{\mu}\,X^{M}\,E_{M}^{\underline{N}}\,\,
\Gamma_{\underline{N}}\,\,,
\label{wvgamma}
\eeq
where $\Gamma_{\underline{N}}$ are constant ten-dimensional Dirac matrices.
Moreover, the pullback of the frame one-forms $e^{\underline{M}}$ is given by
\beq
P[\,e^{\underline{M}}\,]\,=\,E_{N}^{\underline{M}}
\partial_{\mu}\,X^{N}\,d\xi^{\mu}\,\equiv C_{\mu}^{\underline{M}}
\,d\xi^{\mu}\,\,,
\eeq
where, in the last step,  we have defined the coefficients 
$C_{\mu}^{\underline{M}}\equiv E_{N}^{\underline{M}}\partial_{\mu}\,X^{N}$.
Notice that  the induced Dirac matrices $\gamma_{\mu}$ can be expressed in
terms of the constant $\Gamma$'s by means of these same coefficients 
$C_{\mu}^{\underline{M}}$, namely:
\beq
\gamma_{\mu}\,=\,
C_{\mu}^{\underline{M}}\,\Gamma_{\underline{M}}\,\,.
\label{inducedgamma}
\eeq

Let us now decompose the complex spinor $\epsilon$ used up to now in its real and
imaginary parts as
$\epsilon\,=\,\epsilon_1+i\epsilon_2$. We can now arrange the two Majorana-Weyl spinors
$\epsilon_1$ and $\epsilon_2$ as a two-dimensional vector 
$\pmatrix{\epsilon_1\cr\epsilon_2}\,\,$. Acting on these real two-component
spinors, the kappa symmetry matrix of a Dp-brane in the type IIB theory is given by
\cite{swedes}:

\beq
\Gamma_{\kappa}\,=\,{1\over (p+1)!\sqrt{-g}}\,\epsilon^{\mu_1\cdots\mu_{p+1}}\,
(\tau_3)^{{p-3\over 2}}\,i\tau_2\,\otimes\,
\gamma_{\mu_1\cdots\mu_{p+1}}\,\,,
\label{gammakappa}
\eeq
where $g$ is the determinant of the induced metric $g_{\mu\nu}$,
the $\tau_i$ $(i=1,2,3)$ are Pauli matrices that act on the two-dimensional vector 
$\pmatrix{\epsilon_1\cr\epsilon_2}$ and $\gamma_{\mu_1\cdots\mu_{p+1}}$ denotes the 
antisymmetrized product of the induced gamma matrices (\ref{inducedgamma}). In eq.
(\ref{gammakappa}) we have assumed that there are not worldvolume gauge fields on
the Dp-brane and we have taken into account that the Neveu-Schwarz $B$ field is
zero for the Klebanov-Witten background. The absence of worldvolume gauge fields is
consistent with the equations of motion of the probe if there are not source terms
for the worldvolume gauge field in the action. These source terms must be linear in
the gauge field and they can only come from the Wess-Zumino part of the
Dirac-Born-Infeld lagrangian. The former is  responsible for the coupling of the
probe to the Ramond-Ramond fields of the background. In our case we have only one
of such  Ramond-Ramond fields, namely the selfdual five-form $F^{(5)}$. If we
denote by $C^{(4)}$ its potential ($F^{(5)}=dC^{(4)}$), it is clear that the only
term linear in the worldvolume gauge field $A$ in the Wess-Zumino lagrangian is:
\beq
\int F\wedge C^{(4)}\,=\,\int A\wedge F^{(5)}\,\,,
\label{source}
\eeq
where $F=dA$ and we have integrated by parts. In eq. (\ref{source}) it is
understood that  the pullback of the Ramond-Ramond fields to the
worldvolume is being taken. By counting the degree of the form under the integral
in eq. (\ref{source}), it is obvious that such a term can only exist for a
D5-brane and it is zero if the  brane worldvolume
does not capture the  flux of the $F^{(5)}$. As can be easily checked by
inspection, this happens in all the cases studied in this paper except for the
baryon vertex configuration analyzed in appendix A. In this case, the expression
(\ref{gammakappa}) for the kappa symmetry matrix is not valid and one has to use
the more general formula given in ref. \cite{swedes}.

Nevertheless, we could try to find embeddings with non-vanishing worldvolume gauge
fields even when the equations of motion allow to put them to zero. For simplicity,
in this paper we would not try to do this, except for the case studied in
subsection A.3 of appendix A, where a supersymmetric embedding of a D5-brane
with flux of the worldvolume gauge field is obtained.

The supersymmetric BPS configurations of the brane probe are obtained by requiring
the condition:
\beq
\Gamma_{\kappa}\,\epsilon\,=\,\epsilon\,\,,
\label{kappacondition}
\eeq
where $\epsilon$ is a Killing spinor of the background \cite{bbs}. It follows from
eq. (\ref{gammakappa}) that $\Gamma_{\kappa}$ depends on the induced metric and
Dirac matrices, which in turn are determined by the D-brane embedding 
$X^{M}(\xi^{\mu})$. Actually, eq. (\ref{kappacondition}) should be regarded as an
equation whose unknowns are both the embedding $X^{M}(\xi^{\mu})$ and the Killing
spinors $\epsilon$. The number of solutions for $\epsilon$ determines the amount of
background supersymmetry that is preserved by the probe. Notice that we have
written $\Gamma_{\kappa}$ in eq. (\ref{gammakappa}) as a matrix acting on real
two-component spinors, while we have written  the Killing spinors of the background 
in complex notation. However, it is
straightforward to find the following rules to pass from complex to real spinors:
\beq
\epsilon^*\,\leftrightarrow\,\tau_3\,\epsilon\,\,,
\,\,\,\,\,\,\,\,\,\,\,\,\,\,\,\,\,\,\,
i\epsilon^*\,\leftrightarrow\,\tau_1\,\epsilon\,\,,
\,\,\,\,\,\,\,\,\,\,\,\,\,\,\,\,\,\,\,
i\epsilon\,\leftrightarrow\,-i\tau_2\,\epsilon\,\,.
\label{rule}
\eeq
As an example of the application of these rules, notice that the projections 
(\ref{tsprojections}), satisfied by the Killing spinors of the $AdS_5\times T^{1,1}$
background, can be written as:
\beq
\Gamma_{12}\,\otimes\,i\tau_2\,\epsilon\,=\,-
\Gamma_{\hat 1\hat 2}\,\otimes\,i\tau_2\,\epsilon\,=\,\epsilon\,\,.
\label{tprojectionpauli}
\eeq

Let us now discuss the general strategy to solve the kappa symmetry equation 
(\ref{kappacondition}). First of all, notice that, by using the explicit form 
(\ref{inducedgamma}) of the induced Dirac matrices in the expression of
$\Gamma_{\kappa}$ (eq. (\ref{gammakappa})), eq. (\ref{kappacondition}) takes the
form:
\beq
\sum_{i}\,c_i\,\Gamma_{AdS_5}^{(i)}\,
\Gamma_{T^{1,1}}^{(i)}\,\otimes\,
(\tau_3)^{{p-3\over 2}}\,i\tau_2\,\,\epsilon\,=\,\epsilon\,\,,
\label{kappaexplicit}
\eeq
where $\Gamma_{AdS_5}^{(i)}$ ($\Gamma_{T^{1,1}}^{(i)}$) are antisymmetrized products
of constant ten-dimensional Dirac matrices along the $AdS_5$
($T^{1,1}$) directions and the coefficients $c_i$ depend on the embedding 
$X^{M}(\xi^{\mu})$ of the Dp-brane in the $AdS_5\times T^{1,1}$ space. Actually,
due to the relations (\ref{tprojectionpauli}) satisfied by the Killing spinors
$\epsilon$, some of the terms in eq. (\ref{kappaexplicit}) are not independent.
After expressing eq. (\ref{kappaexplicit}) as a sum of independent contributions, we
obtain a new projection for the Killing spinor $\epsilon$. This projection is not,
in general, consistent with the conditions (\ref{tprojectionpauli}) since some of
the matrices  appearing on the left-hand side of eq. (\ref{kappaexplicit}) do not
commute with those appearing in (\ref{tprojectionpauli}). The only way of making
eqs. (\ref{tprojectionpauli}) and (\ref{kappaexplicit}) consistent with each other
is by requiring the vanishing of the coefficients $c_i$ of these non-commuting
matrices, which gives rise to a set of first-order BPS differential equations for
the embedding $X^{M}(\xi^{\mu})$. 

Notice that the kappa symmetry projection of the BPS configurations must be
satisfied at any point of the worldvolume of the brane probe. However, the Killing
spinors $\epsilon$ do depend on the coordinates 
(see eqs. (\ref{adsspinor}) or (\ref{globalspinor})). Thus, it is not obvious at
all that the $\Gamma_{\kappa}\epsilon=\epsilon$ condition can be imposed at all
points of the worldvolume. This fact would be guaranteed if we could recast eq. 
(\ref{kappacondition}) for BPS configurations as an algebraic condition on the
constant spinor $\eta$ of eqs.  (\ref{adsspinor}) or (\ref{globalspinor}). This
algebraic condition on $\eta$ must involve a constant matrix projector and its
fulfillment is generically achieved by imposing some extra conditions to the spinor 
$\epsilon$ (which reduces the amount of supersymmetry preserved by the
configuration) or by restricting appropriately the embedding. For example, when
working on the coordinates (\ref{adspoincare}), one should check whether the kappa
symmetry projector commutes with the matrix $\Gamma_{*}$ of eq.
(\ref{gamma*}). If this is the case, one can consider spinors such as the
$\epsilon_+$'s of eq. (\ref{chiraladsspinor}), which are eigenvectors of 
$\Gamma_{*}$ and, apart from an irrelevant factor depending on the radial
coordinate, are constant. In case we use the parametrization (\ref{globalspinor}),
we should check that, for the BPS embeddings, the kappa symmetry projection
commutes with the matrix multiplying  the spinor $\eta$ on the right-hand side of
eq. (\ref{globalspinor}).

If the BPS differential equations can be solved, one should verify that the
corresponding configuration also solves the equations of motion derived from the
Dirac-Born-Infeld action of the probe. In all the cases analyzed in this paper the
solutions of the BPS equations also solve the equations of motion. Actually, we
will verify that these BPS configurations saturate a bound for the energy, as is
expected for a supersymmetric worldvolume soliton.

\setcounter{equation}{0}
\section{Kappa symmetry for a D3-brane probe}
\medskip
As our first example of D-brane probe in the Klebanov-Witten background, let us
consider a D3-brane. By particularizing eq. (\ref{gammakappa}) to this $p=3$ case,
we obtain that $\Gamma_{\kappa}$ is given by: 
\beq
\Gamma_{\kappa}\,=\,-{i\over 4!\sqrt{-g}}\,\epsilon^{\mu_1\cdots\mu_4}\,
\gamma_{\mu_1\cdots\mu_4}\,\,,
\label{Gammad3}
\eeq
where we have used the dictionary (\ref{rule}) to obtain the expression 
of  $\Gamma_{\kappa}$ acting on complex spinors. 

We will consider several possible configurations with different  number of
dimensions on which the D3-brane is wrapped. Since the  $T^{1,1}$ space is
topologically $S^2\times S^3$, it is natural to consider branes wrapped over three-
and two- cycles. We will study first the case of D3-branes wrapped over a
three-dimensional manifold, where we will find a rich set of BPS configurations.
In appendix A we will allow the D3-brane to be extended along one spacelike
direction of the $AdS_5$ and wrapped over a two-cycle of the  $T^{1,1}$ coset. In
this case we will not be able to find BPS embeddings of the D3-brane probe.
However, we will verify in appendix A that there exist stable, non-supersymmetric,
embeddings of D3-branes wrapping a two-cycle. Actually (see section A.1), these
two-cycles are just the ones obtained in section 4.1, \ie\ those over which a
D5-brane can be wrapped supersymmetrically.

\subsection{D3-branes wrapped on a three-cycle}

Let us use global coordinates  as in eq. (\ref{globalads}) for the $AdS_5$ part of
the metric. We will search for supersymmetric configurations which are pointlike
from the $AdS_5$ point of view and wrap a compact three-manifold within $T^{1,1}$.
Accordingly, let us take the following set of 
worldvolume coordinates:
\beq
\xi^{\mu}\,=\,(\,t,\theta_1,\phi_1,\psi\,)\,\,,
\label{vwcoordinatesd3}
\eeq
and consider embeddings of the type:
\beq
\theta_2\,=\,\theta_2(\theta_1, \phi_1)\,\,,
\,\,\,\,\,\,\,\,\,\,\,\,\,\,\,\,
\phi_2\,=\,\phi_2(\theta_1,\phi_1)\,\,,
\label{ansatzd3}
\eeq
with the radial coordinate $\rho$ and the angles $\alpha^i$  being
constant. For these embeddings $\Gamma_{\kappa}$ in  eq. (\ref{Gammad3}) reduces to:
\beq
\Gamma_{\kappa}\,=\,-iL\,\,{\cosh\rho\over \sqrt{-g}}\,
\Gamma_{t}\,\,\gamma_{\theta_1\phi_1\psi}\,\,.
\label{gammad3s3}
\eeq
The induced gamma matrices along the worldvolume coordinates can be readily obtained
from the general expression (\ref{inducedgamma}). The result is:
\bear
\gamma_{\theta_1}&=&{L\over \sqrt{6}}\,\Big[\,\Gamma_1\,
+\,(\cos\psi\,\partial_{\theta_1}\theta_2
\,+\,\sin\psi\sin\theta_2\partial_{\theta_1}\phi_2\,)
\Gamma_{\hat 1}\,+\,\rc\rc
&&+\,(\,\sin\psi\,\partial_{\theta_1}\theta_2\,-\,
\cos\psi\sin\theta_2\,\partial_{\theta_1}\phi_2\,)\,
\Gamma_{\hat 2}\,\,\Big]\,+\,
{L\over 3}\,\cos\theta_2\,\partial_{\theta_1}\phi_2
\Gamma_{\hat 3}\,\,,\rc\rc
\gamma_{\phi_1}&=&{L\over \sqrt{6}}\,\,\Big[\,\sin\theta_1
\,\Gamma_2\,+\, (\,\sin\theta_2\,
\sin\psi\,\partial_{\phi_1}\phi_2\,+\,\cos\psi\,\partial_{\phi_1}\theta_2
\,)
\Gamma_{\hat 1}\,+\rc\rc
&&+\,(\sin\psi\,\partial_{\phi_1}\theta_2\,-\,
\cos\psi\,\sin\theta_2\,\partial_{\phi_1}\phi_2\,
)\Gamma_{\hat 2}\,\Big]\,
+\,{L\over 3}\,
(\,\cos\theta_1+\,\cos\theta_2\partial_{\phi_1}\phi_2\,)\,
\Gamma_{\hat 3}\,\,,\rc\rc
\gamma_{\psi}&=&{L\over 3}\,\,\Gamma_{\hat 3}\,\,.
\label{inducedgammasd3}
\eear
By using these expressions and the projections (\ref{tsprojections}), it
is easy to verify that:
\beq
{18\over L^3}\,\,
\gamma_{\theta_1\phi_1\psi}\,\epsilon\,=\,ic_1\Gamma_{\hat 3}\epsilon\,+\,
(c_2+ic_3)\,e^{-i\psi}\,\Gamma_{1\hat 2\hat 3}\,\epsilon\,\,,
\label{gammathetaphipsi}
\eeq
with the coefficients $c_1$, $c_2$ and $c_3$ being:
\bear
c_1&=&
\sin\theta_1\,+\,\sin\theta_2\,\Big(\,
\partial_{\theta_1}\theta_2\,\partial_{\phi_1}\phi_2\,-\,
\partial_{\theta_1}\phi_2\,\partial_{\phi_1}\theta_2\,
\Big)\,\,,\rc\rc
c_2&=&
\sin\theta_1\,\partial_{\theta_1}\theta_2\,-\,
\sin\theta_2\partial_{\phi_1}\phi_2\,\,,\rc\rc
c_3&=&
\partial_{\phi_1}\theta_2\,+\,
\sin\theta_1\, \sin\theta_2\,\partial_{\theta_1}\phi_2\,\,.
\label{d3cs}
\eear
Following the general strategy discussed at the end of section 2.2, we have to
ensure that the kappa symmetry projection $\Gamma_{\kappa}\epsilon=\epsilon$ is
compatible with the conditions (\ref{tsprojections}). By inspecting the right-hand
side of eq. (\ref{gammathetaphipsi}) it is fairly obvious that the terms containing
the matrix $\Gamma_{1\hat 2\hat 3}$ would give rise to contributions not compatible
with the projection (\ref{tsprojections}).
Thus, it is clear that to have 
$\Gamma_{\kappa}\,\epsilon=\epsilon$ we must impose the condition
\beq
c_2\,=\,c_3\,=\,0\,\,,
\eeq
which yields the following differential equation for 
$\theta_2(\theta_1,\phi_1)$
and $\phi_2(\theta_1,\phi_1)$:
\bear
\sin\theta_1\,\partial_{\theta_1}\theta_2&=&
\sin\theta_2\,\partial_{\phi_1}\phi_2\,\,,\rc\rc
\partial_{\phi_1}\theta_2&=&-\sin\theta_1\sin\theta_2\,
\partial_{\theta_1}\phi_2\,\,.
\label{bpsd3}
\eear
We will prove below that the first-order equations (\ref{bpsd3}), together with
some extra condition on the Killing spinor $\epsilon$, are enough to ensure that 
$\Gamma_{\kappa}\epsilon\,=\,\epsilon$, \ie\ that our D3-brane probe configuration
preserves some fraction of supersymmetry. For this reason we will refer to 
(\ref{bpsd3}) as the BPS equations of the embedding. It is clear from eq.
(\ref{gammad3s3}) that, in order to compute $\Gamma_{\kappa}$, we need to calculate
the determinant $g$ of the induced metric. From eq. (\ref{inducedmetric}) and the
explicit form (\ref{ansatzd3}) of our ansatz,  it is easy to verify that
the non-vanishing elements of the induced metric are:

\bear
&&g_{\tau\tau}\,=\,-L^2\cosh^2\rho\,\,,\rc\rc
&&g_{\theta_1\theta_1}\,=\,{L^2\over 6}\,\,
\Big[\,1\,+\,(\partial_{\theta_1}\theta_2)^2\,+\,\sin^2\theta_2\,
(\partial_{\theta_1}\phi_2)^2
\Big]\,+\,{L^2\over 9}\,\cos^2\theta_2\,
(\partial_{\theta_1}\phi_2)^2
\,\,,\rc\rc
&&g_{\phi_1\phi_1}\,=\,{L^2\over 6}\,\Big[\,
\sin^2\theta_1\,+\,(\partial_{\phi_1}\theta_2)^2\,+\,
\sin^2\theta_2\,(\partial_{\phi_1}\phi_2)^2\,
\Big]\,+\,{L^2\over 9}\,
\big(\,\cos\theta_1\,+\,\cos\theta_2\,\partial_{\phi_1}\phi_2\,
\big)^2\,\,,\rc\rc
&&g_{\psi\psi}\,=\,{L^2\over 9}\,\,,\rc\rc
&&g_{\phi_1\theta_1}\,=\,{L^2\over 6}\,\Big[\,
\partial_{\theta_1}\theta_2\,\partial_{\phi_1}\theta_2\,+\,
\sin^2\theta_2\,\partial_{\theta_1}\phi_2\,\partial_{\phi_1}\phi_2\,
\Big]\,+\,{L^2\over 9}\,\cos\theta_2\,\partial_{\theta_1}\phi_2\,
(\,\cos\theta_1\,+\,\cos\theta_2\,\partial_{\phi_1}\phi_2\,)\,\,,
\rc\rc
&&g_{\phi_1\psi}\,=\,{L^2\over 9}\,
\big(\,\cos\theta_1\,+\,\cos\theta_2\,\partial_{\phi_1}\phi_2\,\big)
\,\,,\rc\rc
&&g_{\theta_1\psi}\,=\,{L^2\over 9}\,
\cos\theta_2\,\partial_{\theta_1}\,\phi_2\,\,.
\label{inducedmetricd3}
\eear
Let us now define
\bear
\alpha&\equiv&{L^2\over
6}\,\Big[\,1\,+\,(\partial_{\theta_1}\theta_2)^2\,+\,
\sin^2\theta_2\,(\partial_{\theta_1}\phi_2)^2\,\Big]\,\,,\rc\rc
\beta&\equiv&{L^2\over 6}\,\Big[\,
\sin^2\theta_1\,+\,(\partial_{\phi_1}\theta_2)^2\,+\,
\sin^2\theta_2\,(\partial_{\phi_1}\phi_2)^2\,\Big]\,\,,\rc\rc
\gamma&\equiv&{L^2\over 6}\,\Big[\,
\partial_{\theta_1}\theta_2\,\partial_{\phi_1}\theta_2\,+\,
\sin^2\theta_2\,\partial_{\theta_1}\phi_2\,\partial_{\phi_1}\phi_2
\,\Big]\,\,.
\label{alpha}
\eear
From these values one can prove  that
\beq
\sqrt{-g}\,=\,{L^2\cosh\rho\over 3}\,
\sqrt{\alpha\beta\,-\,\gamma^2}\,\,.
\label{detd3}
\eeq
Moreover, if the BPS equations (\ref{bpsd3}) are satisfied, the functions 
$\alpha$, $\beta$ and $\gamma$ take the values:
\beq
\alpha_{|_{BPS}}\,=\,{L^2\over 6\sin\theta_1}\,{c_1}_{|_{BPS}}\,\,,
\,\,\,\,\,\,\,\,\,\,\,\,\,\,
\beta_{|_{BPS}}\,=\,{L^2\sin\theta_1\over 6}\,\,{c_1}_{|_{BPS}}\,\,,
\,\,\,\,\,\,\,\,\,\,\,\,\,\,
\gamma_{|_{BPS}}\,=\,0\,\,,
\eeq
where $c_1$ is written in eq. (\ref{d3cs})
and the determinant of the induced metric is
\beq
\sqrt{-g}_{|_{BPS}}\,=\,{L^4\over 18}\,\cosh\rho\,{c_1}_{|_{BPS}}\,\,.
\eeq

 From this expression of $\sqrt{-g}_{|_{BPS}}$ it is straightforward to
verify that, if the first-order system (\ref{bpsd3}) holds, one has:
\beq
\Gamma_{\kappa}\,\epsilon\,=\,\Gamma_{t}\Gamma_{\hat 3}\,\epsilon\,\,.
\eeq
Thus, the condition $\Gamma_{\kappa}\,\epsilon=\epsilon$ is equivalent to
\beq
\Gamma_{t}\Gamma_{\hat 3}\,\epsilon\,=\,\epsilon\,\,.
\eeq
Let us now plug in this equation the explicit form (\ref{globalspinor}) of
the Killing spinors. Notice that, except for $\Gamma_{\rho}\,\gamma_*$,
$\Gamma_{t}\Gamma_{\hat 3}$ commutes with all matrices appearing on
the right-hand side of eq. (\ref{globalspinor}). Actually, only for
$\rho=0$ the coefficient of $\Gamma_{\rho}\,\gamma_*$ in 
(\ref{globalspinor}) vanishes and, thus, only at this point of $AdS_5$
the equation  $\Gamma_{\kappa}\,\epsilon=\epsilon$ can be
satisfied. In this case, it reduces to the following condition on the
constant spinor $\eta$:
\beq
\Gamma_{t}\Gamma_{\hat 3}\,\eta\,=\,\eta\,\,.
\eeq
Then, in order to have a supersymmetric embedding, we must place our
D3-brane probe at $\rho=0$, \ie\ at the center of the $AdS_5$ space. The
resulting configuration is $1/8$ supersymmetric: it preserves four
Killing spinors of the type (\ref{globalspinor}) with
$\Gamma_{12}\,\eta=-\Gamma_{\hat 1\hat 2}\,\eta\,=\,i\eta$, 
$\,\,\,\Gamma_{t}\Gamma_{\hat 3}\,\eta\,=\,\eta$.

\subsubsection{Integration of the first-order equations}
Let us now integrate the first-order differential equations (\ref{bpsd3}).
Remarkably, this same set of equations has been obtained in ref. \cite{flavoring}
in the study of the supersymmetric embeddings of D5-brane probes in the
Maldacena-N\'u\~nez background \cite{MN}. It was shown in ref. \cite{flavoring} 
that, after a change of variables, the pair of eqs. in (\ref{bpsd3}) can be
converted into the Cauchy-Riemann equations. Indeed, 
let us define two new variables $u_1$ and  $u_2$, related to  
$\theta_1$ and $\theta_2$ as follows:
\beq
u_1\,=\,\log\Big(\tan\,{\theta_1\over 2}\Big)\,\,,
\,\,\,\,\,\,\,\,\,\,\,\,\,\,\,\,\,
u_2\,=\,\log\Big(\tan\,{\theta_2\over 2}\Big)\,\,.
\label{change}
\eeq
Then, it is straightforward to demonstrate that the equations 
(\ref{bpsd3}) can be written as:
\beq
{\partial u_2\over \partial u_1}\,=\,
{\partial \phi_2\over \partial \phi_1}\,\,,
\,\,\,\,\,\,\,\,\,\,\,\,\,\,\,\,\,
{\partial u_2\over \partial \phi_1}\,=\,-
{\partial \phi_2\over \partial u_1}\,\,,
\eeq
\ie\ as the Cauchy-Riemann equations for the variables $(u_1,\phi_1)$ and 
$(u_2,\phi_2)$. Since $u_1,u_2\in (-\infty,+\infty)$ and
$\phi_1,\phi_2\in (0,2\pi)$, the above equations are actually the
Cauchy-Riemann equations in a band. The general integral of these
equations is obtained by requiring that $u_2+i\phi_2$ be an arbitrary
function of the holomorphic variable $u_1+i\phi_1$:
\beq
u_2+i\phi_2\,=\,f(u_1+i\phi_1)\,\,.
\label{generalholo}
\eeq
Let us now consider the particular case in which $u_2+i\phi_2$ depends
linearly on $u_1+i\phi_1$, namely:
\beq
u_2+i\phi_2\,=\,m(u_1+i\phi_1)\,+\,{\rm constant}\,\,,
\label{holo}
\eeq
where $m$ is constant. Let us further assume that $m$ is real and integer. By
equating the imaginary parts of both sides of eq. (\ref{holo}), one gets:
\beq
\phi_2\,=\,m\,\phi_1\,+\,{\rm constant}\,\,.
\label{mwindingphi}
\eeq
Clearly, $m$ can be interpreted as a winding number \cite{flavoring}. Moreover, from
the real part of eq. (\ref{holo}) we immediately obtain $u_2$ as a function
of $u_1$ for this embedding. By using the change of variables of eq.
(\ref{change}) we can convert this $u_2=u_2(u_1)$ function in a relation between
the angles
$\theta_1$ and
$\theta_2$, namely:
\beq
\tan{\theta_2\over 2}\,=\,C\,\Bigg(\,\tan{\theta_1\over 2}\,\Bigg)^m\,\,,
\label{mwindingtheta}
\eeq
with $C$ constant. Following ref. \cite{flavoring} we will call  $m$-winding
embedding to the brane configuration corresponding to eqs. 
(\ref{mwindingphi}) and (\ref{mwindingtheta}).  Notice that for $m=0$ the above
solution reduces to $\theta_2={\rm constant}$, $\phi_2={\rm constant}$. This
zero-winding configuration of the D3-brane is just the one proposed in ref.
\cite{GK} as dual to the dibaryon operators of the $SU(N)\times SU(N)$ gauge
theory. Moreover, when $m=\pm 1$ we have the so-called unit-winding embeddings.
When the constant $C$ in eq. (\ref{mwindingtheta}) is equal to one, it is easy to
find the following form of these unit-winding configurations:
\bear
&&\theta_2=\theta_1\,\,,
\,\,\,\,\,\,\,\,\,\,\,\,\,\,\,\,\,\,\,\,\,\,
\phi_2=\phi_1\,\,,
\,\,\,\,\,\,\,\,\,\,\,\,\,\,\,\,\,\,\,\,\,\,\,\,\,\,
(m=1)\,\,,\rc\rc
&&\theta_2=\pi-\theta_1\,\,,
\,\,\,\,\,\,\,\,\,\,\,\,\,
\phi_2=2\pi-\phi_1\,\,,
\,\,\,\,\,\,\,\,\,\,\,\,\,
(m=-1)\,\,,
\label{unitwinding}
\eear
where we have adjusted appropriately the constant of eq. (\ref{mwindingphi}). Notice
that the two possibilities in (\ref{unitwinding}) correspond to the
two possible identifications of the two $(\theta_1,\phi_1)$ and 
$(\theta_2,\phi_2)$ two-spheres.

\subsubsection{Holomorphic structure}

It is also interesting to write the $m$-winding embeddings just found in
terms of the holomorphic coordinates $z_1,\cdots,z_4$ of the conifold.
Actually, by inspecting eq. (\ref{zratio}), and comparing it with the functions 
$\theta_2=\theta_2(\theta_1)$ and $\phi_2=\phi_2(\phi_1)$ corresponding to a
$m$-winding embedding (eqs. (\ref{mwindingphi}) and (\ref{mwindingtheta})), 
one concludes that the latter can be written,
for example, as\footnote{If the function $f$ in eq. (\ref{generalholo}) satisfies
that $\bar f(z)=f(\bar z)$, then the general solution (\ref{generalholo}) can be
written as $\log{z_1\over z_4}=f\big[\log{z_1\over z_3}\big]$.
}:
\beq
{z_1\over z_4}\,=\,C\,
\Bigg(\,{z_1\over z_3}\,\Bigg)^m\,\,.
\label{d3holomorphic}
\eeq
Thus, the  $m$-winding embeddings of the D3-brane in the $T^{1,1}$ space can be
characterized as the vanishing locus of a polynomial in the $z_i$ coordinates of
$\CC^4$. In order to find this polynomial in its full generality, let us consider
the solutions of the following polynomial equation
\beq
z_1^{m_1}\,z_2^{m_2}\,z_3^{m_3}\,z_4^{m_4}\,=\,{\rm constant}\,\,,
\label{d3pol}
\eeq
where the $m_i's$ are real constants and we will assume that:
\bear
&&m_1+m_2+m_3+m_4=0\,\,,\rc\rc
&&m_1+m_3\not=0\,\,.
\label{d3polconditions}
\eear
By plugging the representation (\ref{holomorphic}) of the $z_i$ coordinates in the
left-hand side of eq. (\ref{d3pol}), one readily proves that, due to the first
condition in eq. (\ref{d3polconditions}), $r$ and $\psi$ are not restricted by eq.
(\ref{d3pol}). Moreover, by looking at the phase of the 
left-hand side of eq. (\ref{d3pol}) one realizes that, if the second condition in 
(\ref{d3polconditions}) holds, $\phi_2$ is related to $\phi_1$ as in eq. 
(\ref{mwindingphi}), with $m$ being given by:
\beq
m\,=\,{m_2+m_3\over m_1+m_3}\,=\,{m_1+m_4\over m_2+m_4}\,\,.
\label{m-winding}
\eeq
Furthermore, from the modulus of eq. (\ref{d3pol}) we easily prove that 
$\theta_2(\theta_1)$ is indeed given by eq. (\ref{mwindingtheta}) with the winding
$m$ displayed in eq. (\ref{m-winding}). As a check of these identifications 
let us notice that, by using the
conifold equation (\ref{conifold}), the embedding (\ref{d3pol}) is
invariant under the change
\beq
(m_1,m_2,m_3,m_4)\rightarrow
(m_1-n,m_2-n,m_3+n,m_4+n)
\label{mchange}
\eeq
for arbitrary $n$. Our relation (\ref{m-winding}) of the winding $m$ and the
exponents
$m_i$  is also invariant under the change (\ref{mchange}).

As particular cases notice that  the zero-winding
embedding can be described by the equation $z_1=Cz_4$, while the unit-winding
solution corresponds to $z_3=C\,z_4$ for $m=1$ and $z_1=C\,z_2$ for $m=-1$. 

For illustrative purposes, let us consider the holomorphic structure of the
solutions (\ref{generalholo}) with a non-linear function $f$. It is easy to see
that, if the function $f$ is not linear, the corresponding holomorphic equation is
non-polynomial. For example, the embedding 
$u_2+i\phi_2\,=\,(u_1+i\phi_1)^2$ corresponds to the equation 
${z_1\over z_4}=\exp[\log^2{z_1\over z_3}]$. Contrary to what happens to the
solutions (\ref{d3pol}) (see subsection 3.1.4), the field theory dual of these
non-polynomial embeddings is completely unclear for us and we will not pursue their
study here.

\subsubsection{Energy bound}

The Dirac-Born-Infeld lagrangian density for the D3-brane probe is given by
\beq
{\cal L}=-\sqrt{-g}\,\,\,,
\label{DBIlagrangian}
\eeq 
where we have taken the D3-brane tension equal to one and the 
 value of $\sqrt{-g}$ for a general embedding of the type 
(\ref{ansatzd3}) has been written in
eq. (\ref{detd3}). We have   checked by explicit calculation that any solution of
the first-order equations (\ref{bpsd3}) also satisfies the Euler-Lagrange equations
of motion derived from the lagrangian  (\ref{DBIlagrangian}). Moreover,
the hamiltonian density for the static configurations we are considering is just
${\cal H}=-{\cal L}$. We are going to prove that ${\cal H}$ satisfies a bound which
is saturated just when the embedding satisfies the BPS equations (\ref{bpsd3}).
To check this fact, let us consider 
arbitrary functions  $\theta_2(\theta_1,\phi_1)$ and
$\phi_2(\theta_1,\phi_1)$. For an embedding at
$\rho=0$, we can write:
\beq
{\cal H}\,=\,{L^2\over 3}\,\,
\sqrt{\alpha\beta-\gamma^2}\,\,,
\eeq
where $\alpha$, $\beta$ and $\gamma$ are given in eq. (\ref{alpha}) and we
have used the value of  $\sqrt{-g}$    given in eq. (\ref{detd3}).  
Let us  now rewrite ${\cal H}$  as
${\cal H}=|{\cal Z}|+{\cal S}$, where 
${\cal Z}\,=\,{L^4\over 18}\,c_1$, with $c_1$ given in the first expression in eq. 
(\ref{d3cs}). It is easily checked that ${\cal Z}$  can be written as a total
derivative
\beq
{\cal Z}\,=\,\partial_{\theta_1}\,{\cal Z}^{\theta_1}\,+\,
\partial_{\phi_1}\,{\cal Z}^{\phi_1}\,\,,
\eeq 
with
\beq
{\cal Z}^{\theta_1}\,=\,-{L^4\over 18}\,\,
(\,\cos\theta_1\,+\,\cos\theta_2\,\partial_{\phi_1}\phi_2)\,\,,
\,\,\,\,\,\,\,\,\,\,\,\,\,\,\,\,\,\,\,\,\,\,\,\,\,\,\,\,\,\,
{\cal Z}^{\phi_1}\,=\,{L^4\over
18}\,\cos\theta_2\,\partial_{\theta_1}\phi_2\,\,.
\eeq
Clearly, ${\cal S}$ is given by:
\beq
{\cal S}\,=\,{L^2\over 3}\,\,
\sqrt{\alpha\beta-\gamma^2}\,-\,{L^4\over 18}\,|c_1|\,\,.
\eeq
Let us now prove that ${\cal S}\ge 0$ for an arbitrary embedding of the type
(\ref{ansatzd3}). Notice that this is equivalent to the following bound
\beq
{\cal H}\,\ge\,|{\cal Z}|\,\,.
\eeq
Moreover , it is easy to verify that the condition ${\cal S}\ge 0$ is
equivalent to:
\beq
\Big(\,\sin\theta_1\,\partial_{\theta_1}\theta_2\,-\,\sin\theta_2
\partial_{\phi_1}\phi_2\,\Big)^2\,+
\Big(\,\partial_{\phi_1}\theta_2\,+\,\sin\theta_1\sin\theta_2\,
\partial_{\theta_1}\phi_2\,\Big)^2
\,\ge\,\,0\,\,,
\label{d3inequality}
\eeq
which is obviously satisfied for arbitrary functions $\theta_2(\theta_1,\phi_1)$ and
$\phi_2(\theta_1,\phi_1)$. Moreover, by inspecting eq. (\ref{d3inequality}) one
easily concludes that the equality in (\ref{d3inequality}) is equivalent to the
first-order BPS equations  (\ref{bpsd3}) and, thus  ${\cal H}=|{\cal Z}|$ for the
BPS embeddings (actually, ${\cal Z}\ge 0$ if the BPS equations are satisfied).

\subsubsection{Field theory dual}
In this subsection we will give some hints on the field theory dual  of a general
D3-brane $m$-winding embedding. First of all, let us try to find the conformal
dimension $\Delta$ of the corresponding operator in the dual field theory. According
to the general AdS/CFT arguments, and taking into account the zero-mode corrections
as in ref. \cite{BHK}, one should have
\beq
\Delta=LM\,\,,
\label{Delta}
\eeq
where $L$ is given in (\ref{KW}) and $M$ is the mass of the wrapped D3-brane. The
latter can be written simply as
\beq
M=T_3\,V_3\,\,,
\eeq
where $V_3$ is the volume of the cycle computed with the induced metric 
(\ref{inducedmetricd3}) and
$T_3$ is the tension of  the D3-brane, given by
\beq
T_3={1\over 8\pi^3(\alpha')^2 g_s}\,\,.
\eeq
Taking into account the results of section 3.1.3, it is not difficult to compute
the value of $V_3$ for  the three-cycle ${\cal C}^{(m)}$ corresponding
to the $m$-winding embedding. The result is:
\beq
V_3\,=\,{8\pi^2 L^3\over 9}\,(1+|m|)\,\,.
\eeq
Now, the mass of the wrapped D3-brane can be readily obtained, namely:
\beq
M={L^3\over 9\pi (\alpha')^2 g_s}\,(1+|m|)\,\,.
\eeq
By plugging this result in eq. (\ref{Delta}), and using the value of $L$ given in
eq. (\ref{KW}), one obtains the following value of the conformal dimension $\Delta$
\beq
\Delta={3\over 4}\,(1+|m|)\,N\,\,.
\label{Dimension}
\eeq
Notice that for $m=0$ we recover from (\ref{Dimension}) the result 
$\Delta={3\over 4}N$ of ref. \cite{GK}, which is the conformal dimension
of  an operator of the form $A^N$, with $A$ being the ${\cal N}=1$
chiral multiplets introduced in section 2 and a double antisymmetrization over the
gauge indices of $A$ is performed in the $N^{{\rm th}}$ power of $A$ (for details
see ref. \cite{GK}). Remember that the conformal dimensions of the
$A$ and
$B$ fields are 
$\Delta(A)=\Delta(B)=3/4$. Thus, in view of the the result (\ref{Dimension}), it is
natural to think that our wrapped D3-branes
for general $m$ correspond to operators with a field content of the form 
\beq
(A^{a}B^{b})^N\,\,\,,
\,\,\,\,\,\,\,\,\,\,\,\,
a+b=1+|m|\,\,.
\label{dualfield}
\eeq
To determine the values of $a$ and $b$ in (\ref{dualfield}) one has to find  the
baryon number of the operator. Recall \cite{KW} that the $U(1)$ baryon number
symmetry acts on the ${\cal N}=1$ matter multiplets as:
\beq
A_i\,\to \,e^{i\alpha}\,A_i\,\,,
\,\,\,\,\,\,\,\,\,\,\,\,
B_i\,\to \,e^{-i\alpha}\,B_i\,\,,
\label{baryonnumber}
\eeq
and, thus, the $A$ ($B$) field has baryon number $+1$($-1$). Notice that the
$z_i$ coordinates are invariant under the transformation (\ref{baryonnumber}) of
the homogeneous coordinates $A_i$ and $B_i$ (see eq. (\ref{homogeneous})). On the
gravity side of the AdS/CFT correspondence, the baryon number (in units of $N$)
can be identified with the third homology class of the three-cycle 
${\cal C}^{(m)}$ over which the D3-brane is wrapped. Indeed, the third homology
group of the $T^{1,1}$ space is 
$H_3(T^{1,1})=\ZZ$. Moreover, the homology class of the cycle can be determined by
representing it as the zero-locus of a polynomial in the $A$ and $B$ coordinates
which transforms homogeneously under the $U(1)$ symmetry (\ref{baryonnumber}).
Actually, the charge of the polynomial under the baryon-number transformations 
(\ref{baryonnumber}) is just the class of the three-cycle ${\cal C}^{(m)}$ in 
$H_3(T^{1,1})=\ZZ$.

It is easy to rewrite the results of section 3.1.2 in
terms of the homogeneous coordinates $A$ and $B$. Indeed, one can prove that the
three-cycle ${\cal C}^{(m)}$ corresponding to the $m$-winding embedding can be
written as
\beq
A_1\,B_2^m\,=\,c\,A_2\,B_1^m\,\,.
\label{homogeneouspol}
\eeq
Notice that changing $m\to -m$ in eq. (\ref{homogeneouspol}) is equivalent to the
exchange $B_1\leftrightarrow B_2$. Therefore, we can always arrange eq.
(\ref{homogeneouspol}) in such a way that the exponents of $A$ and $B$ are
positive\footnote{There is an obvious asymmetry in our equations between the $A$
and $B$ coordinates. The origin of this asymmetry is the particular choice of
worldvolume coordinates we have made in (\ref{vwcoordinatesd3}). If we choose
instead $\xi^{\mu}\,=\,(\,t,\theta_2,\phi_2,\psi\,)$ the role of $A$ and $B$ is
exchanged. Alternatively, the same effect is obtained with the coordinates
(\ref{vwcoordinatesd3}) by changing $m\to 1/m$.  } . Moreover, the polynomial
representing
${\cal C}^{(m)}$ transforms homogeneously under the  symmetry (\ref{baryonnumber}) 
with charge
$1-|m|$, which is just the class of  ${\cal C}^{(m)}$ in  $H_3(T^{1,1})=\ZZ$. One
can confirm this result by computing the integral over ${\cal C}^{(m)}$ of the
(pullback) of the three-form $\omega_3$
\beq
\omega_3\,=\,{1\over 16\pi^2}\,d\psi\,(\,\sin\theta_1d\theta_1d\phi_1\,-\,
\sin\theta_2d\theta_2d\phi_2)\,\,,
\eeq
which has been suitably normalized. 
One can easily check that
\beq
\int_{{\cal C}^{(m)}}\,\omega_3\,=\,1-|m|\,\,,
\eeq
which is the same result as that obtained by representing ${\cal C}^{(m)}$ as in
eq. (\ref{homogeneouspol}). Thus, the baryon number of the dual operator must be
$(1-|m|)N$ and we have a new equation for the exponents $a$ and $b$ in
(\ref{dualfield}), namely $a-b=1-|m|$, which allows to determine the actual values
of $a$ and $b$,  \ie\ $a=1$ and $b=|m|$. Thus, we are led to the conclusion that
the field theory operator dual to our $m$-winding embedding must be of the form:
\beq
\big(\,AB^{|m|}\,\big)^N\,\,.
\label{fieldcontent}
\eeq
We will not attempt to determine here the gauge-invariant index structure of the
operator with the field content (\ref{fieldcontent}) which is dual to the
$m$-winding configurations of the D3-branes. Notice that,   for generic
values of $m$, the absolute value of the baryon number is greater than $N$, whereas
for $m=\pm 1$ it vanishes. This last case resembles that of a giant graviton,
although it is interesting to remember that our unit-winding embeddings are
static, \ie\ time-independent.

\setcounter{equation}{0}
\section{Kappa symmetry for a D5-brane probe}
\medskip
In this section we will explore the possibility of having supersymmetric
configurations of D5-branes which wrap some cycle of the $T^{1,1}$ space. Notice
that, according to the general expression of $\Gamma_{\kappa}$ (eq.
(\ref{gammakappa})) and to the dictionary of eq. (\ref{rule}), one has in this case:
\beq
\Gamma_{\kappa}\,\epsilon\,=\,{i\over 6!\, \sqrt{-g}}\,\,
\epsilon^{\mu_1\cdots\mu_6}\,\gamma_{\mu_1\cdots\mu_6}\,\epsilon^*\,\,.
\label{Gammakappad5}
\eeq
The complex conjugation of the right-hand side of eq. (\ref{Gammakappad5}) will be
of great importance in what follows. Recall that we want 
the D5-brane kappa symmetry projector to be compatible with the 
conditions $\Gamma_{12}\,\epsilon\,=\,i\epsilon$ and 
$\Gamma_{\hat 1\hat 2}\,\epsilon\,=\,-i\epsilon$ of eq. (\ref{tsprojections}).
Since in this D5-brane case the action of $\Gamma_{\kappa}$ on $\epsilon$ involves
the complex conjugation, this compatibility with the conditions
(\ref{tsprojections}) will force us to select embeddings for which the kappa
symmetry projector mixes the two 
$S^2$ spheres, which will allow us to find the differential equations to be
satisfied by the embedding. 

Actually, we will  only be able to carry out successfully this program for the case
of a D5-brane wrapped on a two-cycle. It has been proposed in ref. \cite{GK} that
this kind of configurations represent a domain wall in the gauge theory side.
Indeed, one of such D5-branes is an object of codimension one in $AdS_5$ and, as
argued in ref. \cite{GK}, upon crossing it the gauge theory group changes from
$SU(N)\times SU(N)$ to $SU(N)\times SU(N+1)$. In the next subsection we will
explain in detail how to find such supersymmetric embeddings and we will analyze
some of their properties. In appendix A we will obtain another configuration
preserving the same supersymmetry as the one considered in this section and we will
study the effect of adding flux of the worldvolume gauge fields. In this appendix we
also include a brief account of our unsuccessful attempts to find supersymmetric
embeddings of D5-branes wrapped on a three-cycle. However, we will verify that, by
wrapping the D5-brane over the three-cycles found in section 3.1, one gets stable
non-supersymmetric configurations of the D5-brane probe (see section A.4). 

In appendix A we will also analyze the embeddings in which the D5-brane wraps the
entire $T^{1,1}$, which correspond to the baryon vertex construction in the KW
model. In these configurations the D5-brane worldvolume captures the Ramond-Ramond
flux of the background and the worldvolume gauge field cannot be taken to vanish. We
will study these embeddings both from the point of view of the Dirac-Born-Infeld
action and of the kappa symmetry, and we will conclude that they cannot be
supersymmetric.

\subsection{D5-branes wrapped on a two-cycle}
Let us consider a D5-brane wrapped on a two-cycle of the $T^{1,1}$ space. In order to
preserve supersymmetry in a D3-D5 intersection the two branes must share two spatial
directions. Accordingly, we will place the D5-brane probe at some constant value of
one of the Minkowski coordinates (say $x^3$) and we will extend it along  the radial
direction. Following this discussion, let us take the following set of
worldvolume coordinates
\beq
\xi^{\mu}\,=\,(x^0,x^1,x^2,r,\theta_1,\phi_1)\,\,,
\eeq
and consider embeddings with $x^3$ and $\psi$ constant in which
\beq
\theta_2= \theta_2(\theta_1,\phi_1)\,\,,
\,\,\,\,\,\,\,\,\,\,\,\,\,\,
\phi_2= \phi_2(\theta_1,\phi_1)\,\,.
\eeq
In this case eq. (\ref{Gammakappad5}) takes the form:
\beq
\Gamma_{\kappa}\,\epsilon\,=\,{i\over \sqrt{-g}}\,{r^2\over L^2}\,\,
\Gamma_{x^0x^1x^2r}\,\,\gamma_{\theta_1\phi_1}\,\epsilon^*\,.
\eeq
The induced gamma matrices along the $\theta_1$ and $\phi_1$ directions are 
given by the same equations as in
the D3-brane embeddings of section 3 (see eq. (\ref{inducedgammasd3})). Denoting by
$\psi_0$ the constant value of the $\psi$ coordinate, we obtain after an
straightforward calculation:

\beq
{6\over L^2}\,\,
\gamma_{\theta_1\phi_1}\,\epsilon^*\,\,=\,-ic_1\,\epsilon^*\,+\,(c_2\,-ic_3)
e^{i\psi_0}\,\Gamma_{1\hat 2}\,\epsilon^*\,+
(\,c_4-ic_5)\Gamma_{1\hat 3}\epsilon^*\,+\,
(c_6-ic_7)\,e^{i\psi_0}\,\Gamma_{\hat 1\hat 3}\,\epsilon^*\,\,,
\label{gammakappad5}
\eeq
where the coefficients $c_1$, $c_2$ and $c_3$ are just the same as in the D3-brane 
(eq. (\ref{d3cs})) and
$c_4,\cdots,c_7$ are given by:
\bear
c_4&=&\sqrt{{2\over 3}}\,
\Big[\,\cos\theta_1+\cos\theta_2\partial_{\phi_1}\phi_2\,\Big]\,\,,\rc\rc
c_5&=&-\sqrt{{2\over 3}}\,\,
\sin\theta_1\cos\theta_2\partial_{\theta_1}\phi_2\,\,,\rc\rc
c_6&=&\sqrt{{2\over 3}}\,
\Big[\,\cos\theta_1+\cos\theta_2\partial_{\phi_1}\phi_2\,\Big]\,
\partial_{\theta_1}\theta_2\,\,,\rc\rc
c_7&=&\sqrt{{2\over 3}}\,\Big[\,\cos\theta_1\,+\,
\cos\theta_2\partial_{\phi_1}\phi_2\,\Big]\,
\partial_{\theta_1}\phi_2\,\sin\theta_2\,\,.
\label{c4c7}
\eear

To implement the $\Gamma_{\kappa}\epsilon=\epsilon$ condition one must impose some
differential (BPS) equations which make  some of the $c_i$ coefficients of eq. 
(\ref{gammakappad5}) vanish. The remaining terms give rise to an extra projection
which must commute with the ones already satisfied by the Killing spinors in
order to be compatible with them. By inspecting the different terms on the
right-hand side of eq. (\ref{gammakappad5}), one easily concludes that only the
terms with the 
$\Gamma_{1\hat 2}$ matrix lead to a projection which commutes with the ones corresponding
to the $T^{1,1}$ coset (eq. (\ref{tsprojections})). Therefore, it seems clear that
we must require the vanishing of all the coefficients $c_i$ different from $c_2$ and
$c_3$. Moreover, from the fact that $\Gamma^2_{\kappa}=1$ for any embedding, one
easily proves that
\beq
\sqrt{-g}_{|_{BPS}}\,=\,{r^2\over 6}\,\big|c_2-ic_3\big|_{_{BPS}}\,\,.
\eeq
Then, if $\delta(\theta_1,\phi_1)$ denotes the phase of $c_2-ic_3$, it is clear that, if
the BPS equations hold, the kappa symmetry condition (\ref{kappacondition}) is
equivalent to the projection:
\beq
ie^{i\delta(\theta_1,\phi_1)}\, e^{i\psi_0}\,
\Gamma_{x^0x^1x^2r}\,\,\Gamma_{1\hat 2}\,\epsilon^*\,=\epsilon\,\,.
\label{projdelta}
\eeq
We want to translate the above projection (\ref{projdelta}) into an algebraic condition
involving a constant matrix acting on a constant spinor. It is rather evident by
inspecting eq. (\ref{projdelta}) that this can only be achieved if the phase $\delta$ does
not depend on the worldvolume angles $\theta_1$ and $\phi_1$, which is ensured if
$c_2-ic_3$ is either real or purely imaginary, \ie\ when $c_2$ or $c_3$ is zero. We will
demonstrate later on in this section that by imposing  $c_2=0$ one does not arrive
at a consistent set of equations. Thus, let us consider the case $c_3=0$, \ie\ let
us require that all the coefficients except $c_2$ vanish:
\beq
c_1=c_3=c_4=c_5=c_6=c_7=0\,\,.
\label{cvanishingd5}
\eeq
Notice, first of all, that the condition
$c_5=0$ implies that $\partial_{\theta_1}\phi_2=0$. Substituting this result in the
equation $c_3=0$ one gets that the other crossed derivative $\partial_{\phi_1}\theta_2$
also vanishes  (see eq. (\ref{d3cs})) and, thus, one must have embeddings of the
type:
\beq
\theta_2= \theta_2(\theta_1)\,\,,
\,\,\,\,\,\,\,\,\,\,\,\,\,\,
\phi_2= \phi_2(\phi_1)\,\,.
\label{bpsdependences}
\eeq
For these embeddings $c_7$ is automatically zero and the  conditions 
$c_4=c_6=0$ give rise to the equation
\beq
\cos\theta_1+\cos\theta_2\partial_{\phi_1}\phi_2\,=\,0\,\,,
\label{d5bps1}
\eeq
while the remaining condition $c_1=0$ yields another first-order equation, namely:
\beq
\sin\theta_1\,+\,\sin\theta_2\partial_{\theta_1}\theta_2\partial_{\phi_1}\phi_2\,=\,0\,\,.
\label{d5bps2}
\eeq
Eqs. (\ref{d5bps1}) and (\ref{d5bps2}) are equivalent to the conditions 
(\ref{cvanishingd5}) and are the first-order BPS differential
equations we were looking for in this case.

Let us now try to find the supersymmetry preserved by the BPS configurations. First of all
we notice that
\beq
\sqrt{-g}_{\,\,|_{BPS}}\,=\,{r^2\over 6}\,{|c_2|}_{\,\,_{BPS}}\,\,,
\eeq
and thus, the action of $\Gamma_{\kappa}$ on a Killing spinor $\epsilon$ when the
BPS conditions are satisfied is 
\beq
\Gamma_{\kappa}\,\epsilon_{\,\,|_{BPS}}\,=\,i\,{\rm sign} (c_2)\,e^{i\psi_0}\,
\Gamma_{x^0x^1x^2r}\,\,\Gamma_{1\hat 2}\,\epsilon^*\,\,.
\eeq
Therefore, we must require that:
\beq
i\,{\rm sign} (c_2)\,e^{i\psi_0}\,
\Gamma_{x^0x^1x^2r}\,\,\Gamma_{1\hat 2}\,\epsilon^*\,=\,\epsilon\,\,.
\label{d5projector}
\eeq
We want to convert eq. (\ref{d5projector}) into an algebraic condition on a
constant spinor. With this purpose in mind, let us write the general form of
$\epsilon$ as the sum of the two types of spinors written in eq.
(\ref{chiraladsspinor}), namely:
\beq
\epsilon\,=\,r^{-{1\over 2}}\,\eta_-\,+\,
r^{{1\over 2}}\,\Big(\,{\bar x^3\over L^2}\,
\Gamma_{rx^3}\,\eta_-\,+\,\eta_+\,\Big)\,+
\,{r^{{1\over 2}}\over L^2}\,x^p\,\Gamma_{rx^p}\,\eta_-\,\,,
\label{generalepsilon}
\eeq
where $\bar x^3$ is the constant value of the coordinate $x^3$ in the embedding,
$\eta_{\pm}$ are constant spinors satisfying eq. (\ref{etamasmenos}) and the index
$p$ runs over the set  $\{0,1,2\}$. In eq. (\ref{generalepsilon}) we have
explicitly displayed the dependence of $\epsilon$ on the coordinates $r$ and $x^p$.
By substituting eq.  (\ref{generalepsilon}) on both sides of eq.
(\ref{d5projector}), one can get the conditions that $\eta_{+}$  and
$\eta_{-}$ must satisfy\footnote{A similar analysis, for the D5-brane
configurations in the $AdS_5\times S^5$ background, was performed in ref. \cite{ST}
}. In order to write these conditions, let us define ${\cal
P}$ as the operator that acts on any spinor $\epsilon$ as follows:
\beq
{\cal P}\,\epsilon\,\equiv\,{\rm sign} (c_2) e^{i\psi_0}\,\Gamma_{rx^3}\,
\Gamma_{1\hat 2}\,\epsilon^*\,\,.
\eeq
Then, eq. (\ref{d5projector}) is equivalent to
\bear
&&{\cal P}\,\eta_-\,=\,\eta_-\,\,,\rc\rc
&&(1\,+\,{\cal P}\,)\,\eta_+\,=\,-{2 \bar x^3\over L^2}\,\Gamma_{rx^3}\,\eta_-\,\,.
\label{d5system}
\eear
Since ${\cal P}^2=1$, we can classify the four spinors $\eta_+$ according to their 
${\cal P}$-eigenvalue as:
\beq
{\cal P}\,\eta_+^{(\pm)}\,=\,\pm\eta_+^{(\pm)}\,\,.
\eeq
We can now solve the system (\ref{d5system}) by taking $\eta_-=0$ and taking
$\eta_+$ equal to one of the two
spinors $\eta_+^{(-)}$ of negative ${\cal P}$-eigenvalue. Moreover, there are other
two solutions which correspond to taking a spinor $\eta_+^{(+)}$ of positive 
${\cal P}$-eigenvalue and a
spinor $\eta_-$  related to the former as:
\beq
\eta_{-}\,=\,{L^2\over \bar x^3}\,\Gamma_{r x^3}\,\,\eta_+^{(+)}\,\,.
\label{secondspinord5}
\eeq
Notice that, according to the first equation in (\ref{d5system}), the spinor
$\eta_-$ must have positive ${\cal P}$-eigenvalue, in agreement with eq.
(\ref{secondspinord5}). All together this configuration preserves four
supersymmetries, \ie\ one half of the supersymmetries of the background, 
as expected for a domain wall.

\subsubsection{Integration of the first-order equations}
Let us now integrate the BPS equations (\ref{d5bps1}) and (\ref{d5bps2}). First
of all, notice that eq. (\ref{d5bps1}) can be written as
\beq
{\partial \phi_2\over \partial\phi_1}\,=\,-{\cos\theta_1\over \cos\theta_2}\,=\,
{\rm constant}\,\,,
\label{constancy}
\eeq
where we have already taken into account  the only way in which eq. (\ref{d5bps1})
can be consistent with the dependencies displayed in eq. (\ref{bpsdependences}).
Moreover, by combining eq. (\ref{d5bps2}) with eq. (\ref{d5bps1}), one can
eliminate 
$\partial_{\phi_1}\phi_2$ and obtain the following equation for 
$\partial_{\theta_1}\theta_2$:
\beq
{\partial\theta_2\over \partial\theta_1}\,=\,{\tan\theta_1\over\tan\theta_2}\,\,.
\label{thetaderivative}
\eeq
Eq. (\ref{thetaderivative}) is easily integrated with the result:
\beq
\sin\theta_2=k\sin\theta_1\,\,,
\label{sines}
\eeq 
where $k$ is a constant. Since
$|\cos\theta_2|=\sqrt{1-\sin^2\theta_2}=\sqrt{1-k^2\sin^2\theta_1}$, it follows that
\beq
\Big|{\cos\theta_1\over \cos\theta_2}\,\Big|\,=\,
{|\cos\theta_1|\over  \sqrt{1-k^2\sin^2\theta_1}}\,\,,
\eeq
which is not constant (as required by eq. (\ref{constancy})) unless $k=\pm 1$. It is easy
to conclude that $k$ cannot be equal to $-1$, since in this case $\sin\theta_2$ would be
negative (see eq. (\ref{sines})), which is impossible if $\theta_2\in[0,\pi]$. Thus,
$k=1$ and there are two possibilities $\theta_2=\theta_1$ and 
$\theta_2=\pi-\theta_1$, which correspond respectively to 
$\cos\theta_2/\cos\theta_1=\pm 1$ and 
${\partial\phi_2\over \partial\phi_1}=\mp 1$. Thus, the two solutions are
\bear
&&\theta_2\,=\,\theta_1\,\,,
\,\,\,\,\,\,\,\,\,\,\,\,\,
\phi_2\,=\,2\pi-\phi_1\,\,,
\rc\rc
&&\theta_2\,\,=\,\pi-\theta_1\,\,,
\,\,\,\,\,\,\,\,\,\,\,\,\,
\phi_2\,=\,\phi_1\,\,.
\label{D5embeddings}
\eear
Notice the similarity of (\ref{D5embeddings}) and (\ref{unitwinding}), although the
two solutions are actually very different (see, for example, their different
holomorphic structure). Moreover, it is interesting to point out that
$c_2=2\sin\theta_1$ for the solution with $\theta_2=\theta_1$ while 
$c_2=-2\sin\theta_1$ when  $\theta_2=\pi-\theta_1$. These two solutions correspond to the
two possible signs in the projection (\ref{d5projector}). Moreover, the two-cycles 
(\ref{D5embeddings}) mix the $(\theta_1,\phi_1)$ and $(\theta_2,\phi_2)$
two-spheres, in agreement with the results of \cite{DasMuk}.

To finish this subsection, let us discuss the possibility of requiring the
vanishing of all $c$'s in eq. (\ref{gammakappad5})
except for $c_3$. From the vanishing of  $c_1$,  $c_4$ and  $c_5$ we obtain again eqs. 
(\ref{d5bps1}) and (\ref{d5bps2}). Moreover, from $c_2=0$ we obtain a new equation:
\beq
\partial_{\phi_1}\phi_2\,=\,{\sin\theta_1\over \sin\theta_2}\,\,
\partial_{\theta_1}\,\theta_2\,\,.
\eeq
By combining this new equation with eq. (\ref{d5bps1}), we obtain:
\beq
\partial_{\theta_1}\,\theta_2\,=\,-{\tan\theta_2\over\tan\theta_1}\,\,.
\label{tantheta}
\eeq
It is easy to see that this equation is inconsistent with eq. (\ref{thetaderivative})
(which follows from eqs. (\ref{d5bps1}) and (\ref{d5bps2})). Thus, we conclude that this
way of proceeding does not lead to any new solution of the kappa symmetry condition 
$\Gamma_{\kappa}\,\epsilon=\epsilon$.

\subsubsection{Holomorphic structure}
Let us now write the embeddings just found in terms of the holomorphic coordinates
$z_1,\cdots,z_4$ of the conifold. From the expressions of the ratios of the $z$'s in terms
of the angles $(\theta_i, \phi_i)$ (eq. (\ref{zratio})), it is immediate to realize
that for the $\theta_1=\theta_2$ embedding of eq. (\ref{D5embeddings}) one 
has:
\beq
{z_1\over z_3}\,=\,{z_4\over z_2}\,=\,
{\bar z_1\over \bar z_4}\,=\,{\bar z_3\over \bar z_2}\,\,,
\eeq
which can be written as a single quadratic equation such as:
\beq
z_1\bar z_4\,-\,\bar z_1\,z_3\,=\,0\,\,.
\eeq
(This embedding also satisfies that $|z_3|=|z_4|$). Notice that the equations found are
not holomorphic. This is in correspondence with the fact that the kappa symmetry projector
of a D5 involves a complex conjugation of the spinor. Similarly, for the 
$\theta_2=\pi-\theta_1$ embedding of eq. (\ref{D5embeddings}), one has
\beq
{z_2\over z_3}\,=\,{z_4\over z_1}\,=\,
{\bar z_4\over \bar z_2}\,=\,{\bar z_1\over \bar z_3}\,\,,
\eeq
which, again,  can be recast as a single quadratic equation, which in this case can be
written as:
\beq
z_1\,\bar z_4\,-\, \bar z_2\, z_4\,=\,0\,\,,
\eeq
and one has that $|z_1|=|z_2|$ for this solution.

\subsubsection{Energy bound}
It can be easily verified by explicit calculation that any solution of the type
(\ref{bpsdependences}) of the first-order equations (\ref{d5bps1}) and
(\ref{d5bps2}) is also a solution of the Euler-Lagrange equations derived from the
Dirac-Born-Infeld lagrangian density
${\cal L}=-\sqrt{-g}$. Actually, for a generic configuration 
with $\theta_2= \theta_2(\theta_1)$ and 
$\phi_2= \phi_2(\phi_1)$, the hamiltonian density ${\cal H}=-{\cal L}$ is:
\beq
{\cal H}\,=\,{r^2\over 6}\,\,\sqrt{1\,+\,(\partial_{\theta_1}\theta_2)^2}\,
\sqrt{\sin^2\theta_1\,+\sin^2\theta_2\,(\partial_{\phi_1}\phi_2)^2
+\,{2\over 3}\,(\,\cos\theta_1\,+\,\cos\theta_2\,\partial_{\phi_1}\phi_2\,)^2}\,\,.
\eeq
Let us now show that ${\cal H}$ satisfies a BPS bound of the type
${\cal H}\ge\big|{\cal Z}\big|$, which is saturated precisely when 
(\ref{d5bps1}) and (\ref{d5bps2}) are satisfied. First of all, 
we rewrite ${\cal H}$ as ${\cal H}=\big|{\cal Z}\big|+{\cal S}$, where
\beq
{\cal Z}\,=\,{r^2\over 6}\,\Big(\sin\theta_1\,\partial_{\theta_1}\theta_2\,-\,
\sin\theta_2\,\partial_{\phi_1}\phi_2\,\Big)\,.
\eeq
It can be straightforwardly proven that ${\cal Z}$ can be written as a total
derivative,
\ie
\beq
{\cal Z}\,=\,\partial_{\theta_1}\,{\cal Z}^{\theta_1}\,+\,
\partial_{\phi_1}\,{\cal Z}^{\phi_1}\,\,,
\eeq
and it is not difficult to find the explicit expressions of ${\cal Z}^{\theta_1}$ and
${\cal Z}^{\phi_1}$, namely:
\beq
{\cal Z}^{\theta_1}\,=\,{r^2\over 6}\,\theta_2\sin\theta_1\,\,,
\,\,\,\,\,\,\,\,\,\,\,\,\,\,\,\,\,\,
{\cal Z}^{\phi_1}\,=\,-{r^2\over 6}\,\Big
(\phi_2\sin\theta_2\,+\,\theta_2\phi_1\cos\theta_1\,\Big)
\,\,.
\eeq
Moreover, one can check that ${\cal S}\ge 0$ is equivalent to:
\beq
\Big(\,\sin\theta_1\,+\,\sin\theta_2\,\partial_{\theta_1}\theta_2
\partial_{\phi_1}\phi_2\,\Big)^2\,+\,{2\over 3}\,
\Big(\,1\,+\,(\partial_{\theta_1}\theta_2)^2\,\Big)\,
(\,\cos\theta_1\,+\,\cos\theta_2\,\partial_{\phi_1}\phi_2\,)^2\,
\ge\,0\,\,,
\eeq
which is obviously satisfied and reduces to an equality when the BPS equations 
(\ref{d5bps1}) and (\ref{d5bps2}) are
satisfied. Clearly, in this case the  bound ${\cal H}\ge\big|{\cal Z}\big|$ is
saturated. Notice that ${\cal Z}\ge 0$ for the BPS embedding of
(\ref{D5embeddings}) with $\theta_2=\theta_1$, while  ${\cal Z}\le 0$ when
$\theta_2=\pi-\theta_1$ and $\phi_2=\phi_1$.

\setcounter{equation}{0}
\section{Kappa symmetry for a D7-brane probe}
\medskip
In this section we will try to find supersymmetric embeddings of a D7-brane probe
in the $AdS_5\times T^{1,1}$ geometry. The corresponding kappa symmetry 
 matrix can be obtained from eqs. (\ref{gammakappa}) and (\ref{rule}),
namely:
\beq
\Gamma_{\kappa}\,=\,-{i\over 8!\sqrt{-g}}\,\epsilon^{\mu_1\cdots\mu_8}\,
\gamma_{\mu_1\cdots\mu_8}\,\,.
\label{kappad7}
\eeq
The main interest of studying D7-branes in the $AdS_5\times T^{1,1}$ background
comes from their use as flavor branes, \ie\ as branes whose fluctuations can be
identified with dynamical mesons of the corresponding gauge theory. These flavor
branes must be spacetime filling, \ie\ they must be extended along all the gauge
theory directions. Moreover, their worldvolume should include some holographic,
non-compact, direction. In this section we will determine some supersymmetric
configurations which fulfill these requirements. The study of other embeddings, such
as those in which the D7-brane is wrapped on a three-cycle of $T^{1,1}$, is left for
appendix A. There exists also the possibility of wrapping supersymmetrically the
entire $T^{1,1}$. The corresponding D7-brane embedding preserves two
supersymmetries, as shown in appendix A.

\subsection{Spacetime filling D7-brane}

As explained above, we are interested in D7-brane configurations which extend along
the $x^0\cdots x^3$ coordinates. Notice that in this case the D7-brane probe and
the D3-branes of the background share four dimensions, which is just what is needed
for a supersymmetric D3-D7 intersection. Accordingly, 
let us choose  the following set of worldvolume coordinates 
\beq
\xi^{\mu}\,=\,(x^0,\cdots,x^3,\theta_1,\phi_1,\theta_2,\phi_2)\,\,.
\eeq
The remaining ten-dimensional coordinates $r$ and $\psi$ will be considered as
scalars. Actually, we will restrict ourselves to those configurations in which the
dependence of $r$ and $\psi$ on the worldvolume coordinates is the following:
\beq
\psi\,=\,\psi(\phi_1,\phi_2)\,\,,
\,\,\,\,\,\,\,\,\,\,\,\,
r\,=\,r(\theta_1,\theta_2)\,\,.
\label{d7ansatz}
\eeq
In this case the kappa symmetry matrix (\ref{kappad7}) takes the form:
\beq
\Gamma_{\kappa}\,=\,-i\,{h^{-1}\over \sqrt{-g}}\,\,
\Gamma_{x^0\cdots x^3}\,\,\gamma_{\theta_1\phi_1\theta_2\phi_2}\,\,,
\label{d7kappa}
\eeq
where $h$ is the warp factor of eq. (\ref{KW}). 
The induced gamma matrices along the angular coordinates of the
worldvolume are:
\bear
&&h^{-{1/4}}\,\gamma_{\theta_1}\,=\,{r\over \sqrt{6}}\,\Gamma_1\,+\,
\partial_{\theta_1}r\,\Gamma_r\,\,,\rc\rc
&&h^{-{1/4}}\,\gamma_{\phi_1}\,=\,{r\over \sqrt{6}}\,
\sin\theta_1\,\Gamma_2\,+\,
{r\over 3}\,\,(\cos\theta_1\,+\,\partial_{\phi_1}\psi\,)
\,\Gamma_{\hat 3}\,\,,\rc\rc
&&h^{-{1/4}}\,\gamma_{\theta_2}\,=\,{r\over \sqrt{6}}\,\Big(\,\cos\psi\,
\Gamma_{\hat 1}\,+\,\sin\psi\,\Gamma_{\hat
2}\,\Big)\,+\,\partial_{\theta_2}r\,\Gamma_r\,\,,\rc\rc
&&h^{-{1/4}}\,\gamma_{\phi_2}\,=\,{r\over \sqrt{6}}\,\sin\theta_2\,
\Big(\,\sin\psi\,
\Gamma_{\hat 1}\,-\,\cos\psi\,\Gamma_{\hat 2}\,\Big)\,+\,
{r\over 3}\,(\cos\theta_2\,+\,\partial_{\phi_2}\psi\,)\,
\Gamma_{\hat 3}\,\,.
\label{gammasd7}
\eear
By inspecting the form of the kappa symmetry matrix in eq.
(\ref{d7kappa}), one readily concludes that $\epsilon$ must be a
eigenvector of $\Gamma_*=i\Gamma_{x^0x^1x^2x^3}$. Then, it has to be of
the form $\epsilon_+$ (see eq. (\ref{chiraladsspinor})) and we can
write
\beq
\Gamma_{\kappa}\,\epsilon_+\,=\,-{h^{-1}\over \sqrt{-g}}\,\,
\gamma_{\theta_1\phi_1\theta_2\phi_2}\,\epsilon_+\,\,.
\label{d7gammakappa}
\eeq
Moreover, after taking into account that $\epsilon_+$ has fixed
ten-dimensional chirality
\beq
\Gamma_{x^0x^1x^2x^3}\,\Gamma_{r 12 \hat 1\hat 2\hat 3}\,
\epsilon_+\,=\,-
\epsilon_+\,\,,
\eeq
and using eq. (\ref{tsprojections}), one can easily verify that
\beq
\Gamma_{r\hat 3}\,\epsilon_+\,=\,-i\epsilon_+\,\,.
\label{rhat3}
\eeq
By using these  projection conditions and the explicit form of
the $\gamma$'s (eq. (\ref{gammasd7})), one can prove that
\beq
h^{-1}\,\gamma_{\theta_1\phi_1\theta_2\phi_2}\,\epsilon_+\,=\,
d_1\,\epsilon_+\,+\,ie^{-i\psi}\,d_2\,\Gamma_{\hat 1\hat
3}\,\epsilon_+\,+\, id_3\,\Gamma_{ 1\hat 3}\,\epsilon_+\,+\,
e^{-i\psi}\,d_4\,\Gamma_{ 1\hat 1}\,\epsilon_+\,\,,
\label{gammakappad7}
\eeq
where the coefficients $d_i$ are given by:
\bear
&&d_1\,=\,-{r^4\over 36}\,\sin\theta_1\sin\theta_2\,+\,
{r^3\over 18}\,\Big[\,\sin\theta_1\partial_{\theta_2}r\,
(\,\cos\theta_2\,+\,\partial_{\phi_2}\psi\,)\,+\,
\sin\theta_2\partial_{\theta_1}r\,
(\,\cos\theta_1\,+\,\partial_{\phi_1}\psi\,)\,\Big]\,\,,\rc\rc
&&d_2\,=\,{r^3\over 6\sqrt{6}}\,\sin\theta_1\,
\Big[\,\sin\theta_2\partial_{\theta_2}r\,+\,{r\over 3}\,
(\cos\theta_2\,+\,\partial_{\phi_2}\,\psi\,)\,\Big]\,\,,\rc\rc
&&d_3\,=\,{r^3\over 6\sqrt{6}}\,\sin\theta_2\,
\Big[\,\sin\theta_1\partial_{\theta_1}r\,+\,{r\over 3}\,
(\cos\theta_1\,+\,\partial_{\phi_1}\,\psi\,)\,\Big]\,\,,\rc\rc
&&d_4\,=\,{r^3\over 18}\,\Big[\,\sin\theta_1\
\partial_{\theta_1}r\,(\cos\theta_2\,+\,\partial_{\phi_2}\,\psi\,)\,-\,
\sin\theta_2\
\partial_{\theta_2}r\,(\cos\theta_1\,+\,\partial_{\phi_1}\,\psi\,)
\Big]\,\,.
\eear
Notice that the terms with $d_2$, $d_3$ and $d_4$ on the right-hand side of eq.
(\ref{gammakappad7}) give rise to projections which are not compatible with those
in eq.  (\ref{tsprojections}). Therefore, in order
to have $\Gamma_{\kappa}\,\epsilon_+\,=\,\epsilon_+$,  we impose
\beq
d_2\,=\,d_3\,=\,d_4\,=\,0\,\,.
\eeq
The conditions $d_2=d_3=0$ lead to the following 
differential equations
\beq
\partial_{\theta_1} r\,=\,-{r\over 3}\,
{\cos\theta_1+\partial_{\phi_1}\psi\over \sin\theta_1}\,\,,
\,\,\,\,\,\,\,\,\,\,\,\,\,\,\,\,\,\,\,\,\,\,\,\,
\partial_{\theta_2} r\,=\,-{r\over 3}\,
{\cos\theta_2+\partial_{\phi_2}\psi\over \sin\theta_2}\,\,,
\label{BPS}
\eeq
which, in turn, imply that $d_4=0$. The two differential equations in
(\ref{BPS})  are enough to guarantee the kappa symmetry condition
$\Gamma_{\kappa}\,\epsilon_+\,=\,\epsilon_+$. Actually (see eq.
(\ref{d7gammakappa})),  we have to check
that  $d_1=-\sqrt{-g}$ when the embedding satisfies eq. (\ref{BPS}). This
fact can be easily verified if one uses that, along the angular
directions,  the only non-vanishing elements of the induced metric are:
\bear
&&g_{\theta_i\theta_j}=h^{{1/2}}\,
\big[\,\partial_{\theta_i}r\,\partial_{\theta_j}r\,+\,
{r^2\over 6}\,\delta_{ij}\,\big]\,\,,\rc\rc
&&g_{\phi_i\phi_j}=h^{{1/2}}\,\big[\,
\big(\,\cos\theta_i+\partial_{\phi_i}\psi\,\big)
\big(\,\cos\theta_j+\partial_{\phi_j}\psi\,\big)\,{r^2\over 9}\,+\,
\sin^2\theta_i\,{r^2\over 6}\,\delta_{ij}\,\big]\,\,.
\label{d7inducedmetric}
\eear

From the above analysis it is clear that any Killing spinor of the type 
$\epsilon=\epsilon_+$, with $\epsilon_+$ as in eq. (\ref{chiraladsspinor}),
satisfies the kappa symmetry condition $\Gamma_{\kappa}\epsilon=\epsilon$ if the BPS
equations  (\ref{BPS}) hold. Then, these embeddings preserve the four ordinary 
supersymmetries of the background and, thus, they are $1/8$ supersymmetric.

\subsubsection{Integration of the first-order equations}

The BPS equations (\ref{BPS}) relate the different derivatives of $r$ and $\psi$.
However, notice that according to our ansatz (\ref{d7ansatz}) 
the only dependence on $\phi_1$ and $\phi_2$ in  (\ref{BPS}) comes
from the derivatives of $\psi$. For consistency these derivatives must be
constant, \ie:
\beq
\partial_{\phi_1}\psi\,=\,n_1\,\,,
\,\,\,\,\,\,\,\,\,\,\,\,\,\,\,\,\,\,\,\,\,\,\,\,
\partial_{\phi_2}\psi\,=\,n_2\,\,,
\eeq
where $n_1$ and $n_2$  are two numbers which label the different solutions of the
BPS equations (\ref{BPS}). 
Thus, we can write
\beq
\psi\,=\,n_1\phi_1\,+\,n_2\phi_2\,+\,{\rm constant}\,\,.
\label{winding}
\eeq
We will refer to a solution with given numbers $n_1$ and $n_2$ as a 
$(n_1, n_2)$-winding embedding. Below we will restrict ourselves to the case
in which $n_1$ and $n_2$ are integers. Plugging the function (\ref{winding}) on the
right-hand side of eq. (\ref{BPS}) one gets  an expression for the partial
derivatives of
$r(\theta_1,\theta_2)$ which is easy to integrate. The result is:
\beq
r^3\,=\,{C\over
\Big(\,\sin{\theta_1\over 2}\,\Big)^{n_1+1}\,\,
\Big(\,\cos{\theta_1\over 2}\,\Big)^{1-n_1}\,\,
\Big(\,\sin{\theta_2\over 2}\,\Big)^{n_2+1}\,\,
\Big(\,\cos{\theta_2\over 2}\,\Big)^{1-n_2}}\,\,,
\label{rtheta}
\eeq
where $C$ is a constant of integration. 

Several remarks concerning the function
(\ref{rtheta}) are in order. First of all,
notice that it is impossible to choose $n_1$ and $n_2$ in such a way that $r$ is
constant \footnote{Except when the constant $C$ in eq. (\ref{rtheta}) is equal
to zero. In this case $r=0$, $\sqrt{-g}$ also vanishes and the matrix
$\Gamma_{\kappa}$ is not well-defined.} or even that $r$ depends only on one of the
angles $\theta_i$. Moreover, 
$r(\theta_1,\theta_2)$ is invariant under the change
$\theta_i\rightarrow\pi-\theta_i$ and
$n_i\rightarrow-n_i$, which means that, in the analysis of the function
(\ref{rtheta}),  we can restrict ourselves to the case $n_i\ge 0$. 

It is also
interesting to point out that the function $r(\theta_1,\theta_2)$  always diverges
for some particular values of the angles ($\theta_i=0,\pi$ for $n_i=0$,
$\theta_i=0$ for $n_i\ge 1$ and $\theta_i=\pi$ for $n_i\le -1$). Moreover,
the probe brane reaches the origin $r=0$ of the $AdS_5$ space only when some
of the $n_i$'s is such that $|n_i|\ge 2$. When $n_1$ and $n_2$ are such that
$|n_i|\le 1$ there is a minimum value $r_*$ of the coordinate $r$, which is reached
at some particular values of the $\theta_i$'s. For example, for the 
$(n_1,n_2)=(0,0)$ ($(n_1,n_2)=(1,1)$ ) solution, $r_*^3=4C$ ($r_*^3=C$), and this
value of $r$ is reached when $\theta_1=\theta_2=\pi/2$ ( $\theta_1=\theta_2=\pi$). 
The existence of this minimal value of $r$ is important when one considers these
D7-branes as flavor branes. Indeed, in this case the minimal value of $r$ provides
us of an energy scale, which is naturally identified with the mass of the dynamical
quarks added to the gauge theory. Taking this fact into account, we would be
inclined to think that the above configurations with $n_i=0,\pm 1$ are the most
adequate to be considered as flavor branes for the $AdS_5\times T^{1,1}$
background.

\subsubsection{Holomorphic structure}
\medskip

Let us now prove that the $(n_1,n_2)$-winding embeddings just discussed can be
described by means of a polynomial equation of the type of that written in eq. 
(\ref{d3pol}), where now the exponents $m_i$ must satisfy that
\footnote{For spacetime filling D7-brane embeddings with $m_1+m_2+m_3+m_4=0$, see 
subsection A.6 of appendix A.}:
\beq
m_1+m_2+m_3+m_4\not=0\,\,.
\eeq
Indeed, from the expression of the phase of both sides of eq. (\ref{d3pol}) in
terms of the angular coordinates, one finds that $\psi$ depends on $\phi_1$ and
$\phi_2$ as in eq. (\ref{winding}), with $n_1$ and $n_2$ being given by:
\beq
n_1\,=\,{m_1-m_2-m_3+m_4\over m_1+ m_2+ m_3+ m_4}\,\,,
\,\,\,\,\,\,\,\,\,\,\,\,\,\,
n_2\,=\,{m_1-m_2+m_3-m_4\over m_1+ m_2+ m_3+ m_4}\,\,.
\label{d7midentifications}
\eeq
We can confirm the identification (\ref{d7midentifications}) by extracting 
$r(\theta_1,\theta_2)$ from the modulus of eq. (\ref{d3pol}). Actually, one can
easily demonstrate that the function $r(\theta_1,\theta_2)$ obtained in this way is
just given by the right-hand side of eq. (\ref{rtheta}), with the $n_1$ and $n_2$ of
eq.  (\ref{d7midentifications}). As a further check of these identifications, notice
that $n_1$ and $n_2$ are left invariant under the transformation (\ref{mchange}).

The analysis performed above serves to identify some of our solutions with those
proposed in the literature as flavor branes for this background. In this sense,
notice that the unit-winding
case with $n_1=n_2=1$ corresponds to the  embedding $z_1=C$ proposed in ref.
\cite{Ouyang}, while the zero-winding case with  $n_1=n_2=0$ is the embedding 
$z_1z_2=C$ first considered in ref. \cite{KK} .

\subsubsection {Energy bound}
\medskip
For a D7-brane embedding of  the type (\ref{d7ansatz}), the
Dirac-Born-Infeld lagrangian ${\cal L}\,=\,-\sqrt{-g}$ can be obtained easily from
the elements of the induced metric written in eq. 
(\ref{d7inducedmetric}). We have verified that the 
equations of motion derived from 
${\cal L}$ are satisfied if the BPS first-order equations (\ref{BPS}) are
fulfilled. Actually, as happened with the other supersymmetric embeddings we have
studied, the hamiltonian density ${\cal H}=-{\cal L}$ satisfies a bound of the type
${\cal H}\ge \big| {\cal Z}\big|$, which is saturated just when the BPS equations
(\ref{BPS}) are verified. In order to prove this statement let us consider, as in
our ansatz (\ref{d7ansatz}), arbitrary functions 
$\psi(\phi_1,\phi_2)$ and $r\,=\,r(\theta_1,\theta_2)$.
We define the functions $\Delta_1$ and
$\Delta_2$ as follows
\beq
\Delta_i\,=\,-{r\over 3}\,{\cos\theta_i+\partial_{\phi_i}\psi
\over \sin\theta_i}\,\,,
\,\,\,\,\,\,\,\,\,\,\,\,\,\,
(i=1,2)\,\,.
\eeq
Notice that the BPS equations (\ref{BPS}) 
are just $\partial_{\theta_i}r=\Delta_i$ and  the lagrangian 
${\cal L}\,=\,-\sqrt{-g}$ becomes:
\beq
{\cal L}\,=\,-r^2\sin\theta_1\sin\theta_2
\sqrt{\Big((\partial_{\theta_1}r)^2+(\partial_{\theta_2}r)^2+
{r^2\over 6}\Big)\Big(\Delta_1^2+\Delta_2^2+{r^2\over 6}\Big)}\,\,.
\eeq
Let us now rewrite the hamiltonian density ${\cal H}=-{\cal L}$  as
${\cal H}=\big|{\cal Z}\big|+{\cal S}$, where ${\cal Z}$ is given by:
\beq
{\cal Z}\,=\,r^2\sin\theta_1\sin\theta_2\,\Big(\,{r^2\over 6}\,+\,
\partial_{\theta_1}r\,\Delta_1\,+\,
\partial_{\theta_2}r\,\Delta_2\,\Big)\,\,.
\label{d7Z}
\eeq
When $\psi$ and $r$  are arbitrary functions 
of the type (\ref{d7ansatz}), it is straightforward to prove that 
${\cal Z}$ is a total divergence:
\beq
{\cal Z}\,=\,\partial_{\theta_1}{\cal Z}^{\theta_1}\,+\,
\partial_{\theta_2}{\cal Z}^{\theta_2}\,\,,
\eeq
with ${\cal Z}^{\theta_i}$ being given by:
\beq
{\cal Z}^{\theta_1}\,=\,-{r^4\over 12}\,(\cos\theta_1+\partial_{\phi_1}\psi)
\sin\theta_2\,\,,
\,\,\,\,\,\,\,\,\,\,\,\,\,\,\,\,\,\,\,\,\,\,\,\,\,\,\,\,
{\cal Z}^{\theta_2}\,=\,-{r^4\over 12}\,(\cos\theta_2+\partial_{\phi_2}\psi)
\sin\theta_1\,\,.
\eeq
Moreover, the function ${\cal S}={\cal H}-\big|{\cal Z}\big|$ is non-negative. 
Actually, the condition ${\cal S}\ge 0$ is equivalent to:
\beq
{r^2\over 6}\,\Big(\partial_{\theta_1}r\,-\,\Delta_1\Big)^2\,+\,
{r^2\over 6}\,\Big(\partial_{\theta_2}r\,-\,\Delta_2\Big)^2\,+\,
\Big(\partial_{\theta_1}r\,\Delta_2\,-\,\partial_{\theta_2}r\,\Delta_1
\Big)^2\,\ge\,0\,\,,
\eeq
which is obviously always satisfied and reduces to an equality when 
$\partial_{\theta_i}r=\Delta_i$. Thus, ${\cal H}\ge \big| {\cal Z}\big|$ and, as
previously claimed, the BPS conditions (\ref{BPS}) saturate the bound. It is also
clear from the expression of ${\cal Z}$ in (\ref{d7Z}) that  ${\cal Z}_{|BPS}\ge
0$. 

\section{Summary and conclusions}

Let us summarize our main results. We have used kappa symmetry to explore in a
systematic way the supersymmetric embeddings of  D-brane probes in the
$AdS_5\times T^{1,1}$ geometry. Our method is based on a detailed knowledge of the
Killing spinors of the background and allows to determine the explicit form
of the D-brane embedding, as well as the fraction of supersymmetry preserved by the
different configurations. Generically, the supersymmetric
embeddings are obtained by integrating a system of first-order BPS differential
equations. We have checked in all cases that the solutions of these BPS equations
also solve the equations of motion derived from the Dirac-Born-Infeld action of the
brane probe. Actually, we have verified that our embeddings saturate an energy
bound, as it is expected to occur for a worldvolume soliton.

In the case of a D3-brane we have found a family of
three-cycles, which generalizes the ones used in ref. \cite{GK} to construct the
duals of the dibaryon operators, and we have determined the field content of the
dual operator for our cycles.  We  have also been able to find explicitly the
two-cycles over which one must wrap a D5-brane in order to preserve some fraction of
supersymmetry, which should correspond to domain walls in the field theory dual.
The analysis of spacetime filling configurations of D7-brane probes led us to
determine a two-parameter family of supersymmetric embeddings, some of those
having the right properties to be considered as flavor branes for the
Klebanov-Witten model. Moreover, as shown in appendix A, the D7-brane can wrap
completely the $T^{1,1}$ coset and preserve two supersymmetries.
We have also shown (see appendix A) that there exist stable
non-supersymmetric configurations of a D3-brane (D5--brane) wrapped over a
two-cycle (three-cycle). The baryon vertex construction (a D5-brane wrapping the 
$T^{1,1}$ space) is also studied in appendix A and we conclude that the
corresponding configuration is not supersymmetric.
All the supersymmetric embeddings we have found can be
described by means of a simple polynomial equation in terms of the holomorphic
coordinates of the conifold. Notice, however, that our method does not rely on this
fact and can be applied to other backgrounds, as in ref. \cite{MN}, for which the
algebraic-geometric techniques are not available. Nevertheless, one might wonder
if, for the background studied here, it would not be more appropriate to use the
holomorphic coordinates from the beginning, instead of the angular and radial
variables that we have employed. To clarify this point, let us recall that the
$z$ coordinates are not independent, since they satisfy the constraint
(\ref{conifold}) and, therefore, their use as worldvolume coordinates would be
rather cumbersome.

Let us now discuss some possible extensions of our work. We could study the
fluctuations of the brane probe around the static configurations we have found. In
the case of the D3-brane, this has already been done in ref. \cite{BHK} for the
zero-winding embedding. Moreover, the fluctuations of the spacetime filling
D7-branes would allow us to extract the spectrum of dynamical mesons of the theory,
as was done in refs. \cite{KMMW,Sonnen,Johana,flavoring} for different cases.
Another interesting future research line would be the application of our
methodology to other backgrounds. Recall that the starting point of our formalism
is a representation of the Killing spinors in a basis in which they become
independent of the compact coordinates. For the Klebanov-Strassler background such
a representation is obtained in appendix B. It would also be interesting to study
the supersymmetric configurations of M2 and M5 brane probes in some backgrounds of
eleven-dimensional supergravity. The analogue of the case studied here would be
considering a manifold of the type $AdS_4\times X_7$, where $X_7$ is a
seven-dimensional Einstein space \cite{Kirsch}. Work along these lines is in
progress and we hope to report on it in a near future.

\medskip
\section*{Acknowledgments}
\medskip
We are grateful to Jos\'e D. Edelstein, Sean A.  Hartnoll, Carlos N\'u\~nez,
Angel Paredes, Ruben Portugues and Martin Schvellinger for useful discussions and
encouragement.   This work  is supported in part by MCyT, FEDER and Xunta de
Galicia under grant  BFM2002-03881 and by  the EC Commission under the FP5 grant
HPRN-CT-2002-00325.

\vskip 1cm
\renewcommand{\theequation}{\rm{A}.\arabic{equation}}
\setcounter{equation}{0}
\medskip
\appendix
\section{Some other possibilities }
\medskip
In this appendix we explore some possible configurations of brane probes which have
not been analyzed in the main text. The first case we will study is a D3-brane
extended along one direction in the $AdS_5$ space and wrapped along two directions
of the $T^{1,1}$ coset. We will verify that such ``fat" string configurations are
not supersymmetric.  However, we will show that there exist stable
non-supersymmetric embeddings in which the D3-brane wraps the two-cycle found
in section 4.1.

We will then consider D5-branes
wrapped over two- and three-dimensional submanifolds of the $T^{1,1}$ space. In the
first case we will find a supersymmetric configuration not included among those
analyzed in section 4 and we will show that one can add flux of the worldvolume
gauge field and preserve the same amount of supersymmetry.
On the contrary, we will not be able to find any
supersymmetric solution for a D5-brane probe wrapped on a three-cycle.
Nevertheless, one can wrap the D5-brane on the three-cycles found in section 3.1.
We will show that these embeddings are stable non-supersymmetric solutions of the
equations of motion. We will also analyze the baryon vertex configuration, which
corresponds to a D5-brane wrapping the entire $T^{1,1}$ space. In this case the
worldvolume gauge fields cannot be taken to vanish and, despite its similarity
with the baryon vertex construction in $AdS_5\times S^5$, we will conclude that
supersymmetry is broken completely.

We will study again the spacetime filling configurations of D7-branes by
using an alternative set of worldvolume coordinates which now will include the
radial variable $r$. As a result of this analysis, we will find a class of
embeddings in which the D7-brane extends infinitely in $r$ and wraps the
three-cycles found in section 3.1. These configurations preserve the same
supersymmetries as those found in section 5.1 and can be represented by a polynomial
equation of the same type in the holomorphic coordinates of the conifold. Finally,
we will show that it is possible to wrap a D7-brane over the full $T^{1,1}$ coset
and preserve some supersymmetry.

\subsection{D3-branes wrapped on a two-cycle}
Let us consider now a D3-brane which shares one direction (say $x^1$) with the
D3-branes of the background and wraps a two-dimensional cycle of the
compact $T^{1,1}$ space. Such an object would be one-dimensional from the gauge
theory perspective, \ie\ it would look like as an inflated string. To describe such
a configuration we will take the following set of worldvolume coordinates
$\xi^{\mu}=(x^0,x^1,\theta_1,\phi_1)$ and we will look for embeddings with $x^2$,
$x^3$, $r$ and $\psi$ constant and $\theta_2=\theta_2(\theta_1,\phi_1)$ and 
$\phi_2=\phi_2(\theta_1,\phi_1)$. In this case the kappa symmetry matrix
$\Gamma_{\kappa}$ acts on a Killing spinor $\epsilon$ as:
\beq
\Gamma_{\kappa}\,\epsilon\,=\,-{i\over \sqrt{-g}}\,{r^2\over L^2}\,
\Gamma_{x^0x^1}\,\gamma_{\theta_1\phi_1}\,\epsilon\,\,.
\eeq
The induced matrices $\gamma_{\theta_1}$ and $\gamma_{\phi_1}$ are the same as
those written in eq. (\ref{inducedgammasd3}) and the action of
$\gamma_{\theta_1\phi_1}$ on 
$\epsilon$ can be obtained as in eq. (\ref{gammakappad5}):
\beq
{6\over L^2}\,\,
\gamma_{\theta_1\phi_1}\,\epsilon\,\,=\,ic_1\,\epsilon\,+\,(c_2\,+ic_3)
e^{-i\psi_0}\,\Gamma_{1\hat 2}\,\epsilon\,+
(\,c_4+ic_5)\Gamma_{1\hat 3}\epsilon\,+\,
(c_6+ic_7)\,e^{-i\psi_0}\,\Gamma_{\hat 1\hat 3}\,\epsilon\,\,,
\label{fatstring}
\eeq
where $c_1$, $c_2$ and $c_3$ are given in eq. (\ref{d3cs}), while $c_4,\cdots, 
c_7$ are displayed in eq. (\ref{c4c7}).

The only projection compatible with the $T^{1,1}$ conditions (\ref{tsprojections})
is the one originated from the first term on the right-hand side of 
eq. (\ref{fatstring}). Accordingly, let us require that $c_2=c_3=\cdots=c_7=0$. From
$c_2=c_3=0$ we obtain the Cauchy-Riemann equations (\ref{bpsd3}). Moreover, from
$c_5=0$ it follows that $\phi_2=\phi_2(\phi_1)$ which ensures that
$c_7$ vanishes, while by using this result on the second equation in (\ref{bpsd3})
one concludes that $\theta_2=\theta_2(\theta_1)$. On the other hand, the conditions
$c_4=c_6=0$ lead to eq. (\ref{constancy}), \ie\ to
$\partial_{\phi_1}\phi_2=-\cos\theta_1/\cos\theta_2={\rm constant}$. By combining
this result with the first equation in (\ref{bpsd3}), one gets the same result  as
in eq. (\ref{tantheta}),
namely $\partial_{\theta_1}\theta_2=-\tan\theta_2/\tan\theta_1$, which can be easily
integrated with the result $\sin\theta_2\sin\theta_1=k$, with $k$ being a constant
which necessarily must satisfy $k\le 1$. It is not difficult to show that this
solution is inconsistent since, for example, $\cos\theta_1/\cos\theta_2$ cannot be
constant. 

If we give up the requirement of supersymmetry, it is not difficult to find stable
solutions of the equations of motion. Indeed, apart from  an irrelevant global
factor, the lagrangian of the D3-brane probe considered here is given by the same
expression as the one obtained in section 4.1 for  a D5-brane wrapping a
two-cycle. It follows from our analysis of section 4.1 that any solution of eqs. 
(\ref{d5bps1}) and (\ref{d5bps2}) also solves the Euler-Lagrange equations of
motion. Recall that the solutions of eqs. (\ref{d5bps1}) and (\ref{d5bps2}) have
been written in eq. (\ref{D5embeddings}). Thus, the functions (\ref{D5embeddings})
are also a solution of the D3-brane equations of motion. Moreover, from the
equivalence between the D3- and D5-brane hamiltonians and the results of section
4.1.3, one can also establish a bound for the D3-brane energy, which is saturated
for the configurations (\ref{D5embeddings}). This fact ensures that these
embeddings, although they break supersymmetry completely, are stable.

\subsection{More D5-branes wrapped on a two-cycle}

Let us consider a D5-brane wrapped on a two-cycle of the $T^{1,1}$ space and let us
take the following set of worldvolume coordinates
$
\xi^{\mu}\,=\,(x^0,x^1,x^2,r,\theta_1,\theta_2)\,\,.
$
We shall consider embeddings with $x^3$ and $\psi$ constant in which
$
\phi_1= \phi_1(\theta_1,\theta_2)\,\,,
\phi_2= \phi_2(\theta_1,\theta_2)\,\,.
$
Particularizing eq. (\ref{Gammakappad5}) to this case, we arrive at:
\beq
\Gamma_{\kappa}\,\epsilon\,=\,{i\over \sqrt{-g}}\,{r^2\over L^2}\,\,
\Gamma_{x^0x^1x^2r}\,\,\gamma_{\theta_1\theta_2}\,\epsilon^*\,.
\eeq
Denoting by $\psi_0$ the constant value of $\psi$, we obtain the following value of the
induced gamma matrices:
\bear
\gamma_{\theta_1}&=&{L\over \sqrt{6}}\Big[
\Gamma_1\,+\,\sin\theta_1\partial_{\theta_1}\phi_1\,\Gamma_2\,+\,
\sin\theta_2\partial_{\theta_1}\phi_2\,
\big(\,\sin\psi_0\,\Gamma_{\hat 1}\,-\,\cos\psi_0\Gamma_{\hat 2}\,\big)\,\Big]\,+\rc\rc
&&+\,{L\over 3}\,
\Big[\,\cos\theta_1\partial_{\theta_1}\phi_1+\,\cos\theta_2\partial_{\theta_1}\phi_2
\,\Big]\,\Gamma_{\hat 3}\,\,,\rc\rc
\gamma_{\theta_2}&=&{L\over \sqrt{6}}
\Big[\,\big(\cos\psi_0+\sin\psi_0\sin\theta_2\partial_{\theta_2}\phi_2\big)
\,\Gamma_{\hat 1}\,+\,
\big(\sin\psi_0-\cos\psi_0\sin\theta_2\partial_{\theta_2}\phi_2\big)\,\Gamma_{\hat 2}\,
+\,\sin\theta_1\partial_{\theta_2}\phi_1\,\Gamma_{ 2}\,\Big]\,+\	\rc\rc
&&\,+\,{L\over 3}\,
\Big[\,\cos\theta_1\partial_{\theta_2}\phi_1+\,\cos\theta_2\partial_{\theta_2}\phi_2
\,\Big]\,\Gamma_{\hat 3}\,\,.
\eear
After an straightforward calculation, one can demonstrate that:
\beq
\gamma_{\theta_1\theta_2}\,\epsilon^*\,\,=\,if_1\,\epsilon^*\,+\,(f_2+if_3)\,
e^{i\psi_0}\,\Gamma_{1\hat 1}\,\epsilon^*\,+\,(f_4+if_5)\Gamma_{1\hat
3}\epsilon^*\,+\, (f_6+if_7)\,e^{i\psi_0}\,\Gamma_{\hat 1\hat 3}\,\epsilon^*\,\,,
\eeq
with $f_1,\cdots,f_7$ being given by:
\bear
f_1&=&{L^2\over 6}\,\,\Big[
\sin\theta_2\partial_{\theta_1}\phi_2\,-\,\sin\theta_1\partial_{\theta_2}\phi_1
\,\Big]\,\,,    \rc\rc
f_2&=&{L^2\over 6}\,\Big[\,1\,+\,\sin\theta_1\sin\theta_2\,\big(\,
\partial_{\theta_2}\phi_1\partial_{\theta_1}\phi_2-
\partial_{\theta_1}\phi_1\partial_{\theta_2}\phi_2\,\big)\,\Big]\,\,,\rc\rc
f_3&=&-{L^2\over 6}\,\Big[\,\sin\theta_1\partial_{\theta_1}\phi_1+
\sin\theta_2\partial_{\theta_2}\phi_2\,\Big]\,\,,\rc\rc
f_4&=&{L^2\over 3\sqrt{6}}\,\Lambda_2\,\,,\rc\rc
f_5&=&{L^2\over 3\sqrt{6}}\,
\sin\theta_1\,\Big[\,
\partial_{\theta_2}\,\phi_1\,\Lambda_1\,-\,
\partial_{\theta_1}\phi_1\,\Lambda_2\,\Big]\,\,,\rc\rc
f_6&=&-{L^2\over 3\sqrt{6}}\,\Lambda_1\,\,,\rc\rc
f_7&=&{L^2\over 3\sqrt{6}}\,
\sin\theta_2\,\Big[\,
\partial_{\theta_2}\,\phi_2\,\Lambda_1\,-\,
\partial_{\theta_1}\phi_2\,\Lambda_2\,\Big]\,\,,
\label{fs}
\eear
and we have introduced the quantities $\Lambda_1$ and $\Lambda_2$, defined as:
\beq
\Lambda_i\equiv\cos\theta_1\partial_{\theta_i}\phi_1\,+\,
\cos\theta_2\partial_{\theta_i}\phi_2\,\,,
\,\,\,\,\,\,\,\,\,\,\,\,\,\,\,\,\,\,\,\,\,\,
(i=1,2)\,\,.
\eeq
In order to have a projection compatible with that of eq. (\ref{tsprojections}), we
shall require
\beq
f_1=f_4=f_5=f_6=f_7=0\,\,.
\eeq
Taking into account that the vanishing of $f_4$, $f_5$, $f_6$ and $f_7$
is equivalent to the conditions $\Lambda_1=\Lambda_2=0$, we arrive at the
following system of first-order differential equations:
\bear
\sin\theta_1\partial_{\theta_2}\phi_1&=&\sin\theta_2\partial_{\theta_1}\phi_2\,\,,\rc\rc
\cos\theta_1\partial_{\theta_i}\phi_1&=&-\cos\theta_2\partial_{\theta_i}\phi_2\,\,,
\,\,\,\,\,\,\,\,\,\,\,\,\,\,(i=1,2)\,\,.
\label{newd5system}
\eear
A solution of (\ref{newd5system})  can be found by the method of separation of
variables. The result is:
\bear
\phi_1&=&A\,\,\Bigg({\cos\theta_2\over\cos\theta_1}\Bigg)^{\alpha}\,+\,\phi_1^{0}\,\,,\rc\rc
\phi_2&=&{\alpha\over 1-\alpha}\,
A\,\,\Bigg({\cos\theta_1\over\cos\theta_2}\Bigg)^{1-\alpha}\,+\,\phi_2^{0}\,\,,
\label{solution}
\eear
where $A$, $\alpha$, $\phi_1^{0}$ and $\phi_2^{0}$ are constants. Notice that, when
the constant $A$ (or $\alpha$) vanishes, the above solution reduces to that in which
$\phi_1$ and $\phi_2$ are both constant. Moreover, in agreement with our
discussion of sect. 4, the phase of $f_2+if_3$ should be constant. By plugging the 
solution written in (\ref{solution}) into the expressions of $f_2$ and $f_3$ (eq. 
(\ref{fs})), 
it is easy to convince oneself that this only
happens if $A=0$, \ie\ for the solution with $\phi_1$ and $\phi_2$  constant.
Therefore, this is the only admissible solution. Actually, one can easily check that
it satisfies the equations of motion and  preserves the same four supersymmetries as
in eq. (\ref{d5projector}) with ${\rm sign}(c_2)$ changed by ${\rm sign}(f_2)=+1$. 

\subsection{D5-branes wrapped on a two-cycle with flux }

We want now  to analyze the effect of adding flux of the worldvolume gauge field to
the configurations studied in section 4. Accordingly, 
let us switch on  $q$ units of worldvolume flux along the angular directions
$\theta_1$ and $\phi_1$. The corresponding field strength is
\beq
F_{\theta_1\phi_1}\,=\,q\sin\theta_1\,\,.
\label{vwflux}
\eeq
When $q\not =0$, the Wess-Zumino term of the Dirac-Born-Infeld action does not
vanish anymore. Indeed, the pullback $P[C^{(4)}]$ of the Ramond-Ramond four-form
potential has a component of the form
\beq
P[C^{(4)}]_{x^0 x^1 x^2 r}\,=\,h^{-1}\,{dx^3\over dr}\,\,,
\eeq
and the corresponding term in the action is:
\beq
{\cal L}_{WZ}\,=\,-q\,h^{-1}\,x'\,\sin\theta_1\,\,,
\eeq
where we have denoted $x^3$ simply as $x$ and the prime means derivation with respect to
$r$. The existence of this term implies that $x^3$ cannot be taken to be
independent of $r$ if $q$ is non-vanishing. In order to find this dependence, let
us write the lagrangian density for the same angular embedding as in the zero-flux
case (eq. (\ref{D5embeddings})) and for an arbitrary function $x(r)$. One gets:
\beq
{\cal L}\,=\,\sin\theta_1\,\Big[\,-\bigg(\,{L^4\over 9}\,+\,q^2\,\bigg)^{{1\over 2}}\,
h^{-{1\over 2}}\,\,\big(1\,+\,h^{-1}\,(x')^2\,)^{{1\over 2}}\,-\,q\,h^{-1}\,x'\,
\Big]\,\,.
\eeq
The equation of motion for $x$ derived from ${\cal L}$ implies
\beq
\bigg(\,{L^4\over 9}\,+\,q^2\,\bigg)^{{1\over 2}}\,\,\,
{h^{-{3\over 2}}\,x'\over \big(1\,+\,h^{-1}\,(x')^2\,)^{{1\over 2}}}\,+\,
q\,h^{-1}\,=\,{\rm constant}\,\,.
\label{eomflux}
\eeq
Let us take the constant on the right-hand side of eq. (\ref{eomflux}) equal to
zero. Then,   equation (\ref{eomflux}) implies
\beq
(x')^2\,=\,{9q^2\over r^4}\,\,.
\eeq
Let us consider the solution of the equation of motion with
\beq
x'\,=\,-{3q\over r^2}\,\,.
\label{xprime}
\eeq
which, after integration, becomes:
\beq
x\,=\,\bar x^3\,+\,{3q\over r}\,\,,
\label{xr}
\eeq
with $\bar x^3$ constant. We are now going to prove that this solution of the
equations of motion is supersymmetric. Notice that the expression of the
kappa symmetry matrix $\Gamma_{\kappa}$ differs from that written in eq.
(\ref{gammakappa}), due to the non-zero value of the worldvolume gauge field.
Actually, from the general expression of $\Gamma_{\kappa}$ given in ref.
\cite{swedes}, one can easily prove that, for the case at hand, one has
\beq
\Gamma_{\kappa}\,\epsilon\,=\,{i\over 
\sqrt{-\det (g+F)}}\,\,{r^3\over L^3}\,\Gamma_{x^0x^1x^2}\,
\Bigg[\,\gamma_r\,\gamma_{\theta_1\phi_1}\,\epsilon^*\,
-\,\gamma_r\,F_{\theta_1\phi_1}\,\epsilon\,\Bigg]\,\,.
\eeq
Notice that $\gamma_r$ is given by:
\beq
\gamma_r\,=\,{L\over r}\,\big(\,\Gamma_r\,+\,{r^2\over L^2}\,x'\,
\Gamma_{x^3}\,\big)\,\,,
\eeq
and for the angular embeddings of eq. (\ref{D5embeddings}) we have proved in
section 4 that 
\beq
\gamma_{\theta_1\phi_1}\,\epsilon^*\,=\,{\rm sign} (c_2)\,{L^2\over 3}\sin\theta_1\,
e^{i\psi_0}\,\Gamma_{1\hat 2}\,\epsilon^*\,\,.
\eeq
Taking $x'$ and $F_{\theta_1\phi_1}$ as given in eqs. (\ref{xprime}) and
(\ref{vwflux}) respectively, one gets:
\beq
\Gamma_{\kappa}\,\epsilon\,=\,{i\over 1+{9q^2\over L^4}}\,\,
\Gamma_{x^0x^1x^2 r}\Big[\,
{\rm sign} (c_2)\,e^{i\psi_0}\,\Gamma_{1\hat 2}\,\epsilon^*\,-\,
{3q\over L^2}\,{\rm sign} (c_2)\,e^{i\psi_0}\,\Gamma_{r x^3}\,\Gamma_{1\hat
2}\,\epsilon^* -{3q\over L^2}\,\epsilon\,+\,{9q^2\over
L^4}\,\Gamma_{rx^3}\,\epsilon\,\Big]\,\,.
\eeq
Moreover, by plugging in  eq. (\ref{adsspinor}) the explicit dependence of $x$ on
$r$ (eq. (\ref{xr})), one gets that the Killing spinor $\epsilon$ evaluated on the
worldvolume can be written as:
\beq
\epsilon\,=\,r^{-{1\over 2}}\,\Big(\,1\,+\,{3 q\over L^2}\,\Gamma_{rx^3}\Big)
\eta_-\,+\,
r^{{1\over 2}}\,\Big(\,{\bar x^3\over L^2}\,
\Gamma_{rx^3}\,\eta_-\,+\,\eta_+\,\Big)\,+
\,{r^{{1\over 2}}\over L^2}\,x^p\,\Gamma_{rx^p}\,\eta_-\,\,,
\label{epsilonflux}
\eeq
where the constant spinors $\eta_{\pm}$ have been defined in eq.
(\ref{etamasmenos}). By using the expression of $\epsilon$ given in eq.
(\ref{epsilonflux}) in the equation $\Gamma_{\kappa}\epsilon\,=\,\epsilon$, one
finds that, remarkably, the kappa symmetry condition is verified if 
$\eta_+$ and $\eta_-$ satisfy the system (\ref{d5system}). Therefore, this
configuration preserves the same four supersymmetries as in the zero flux case
studied in section 4.

\subsection{D5-branes wrapped on a three-cycle}

Let us now explore the possibility of having D5-brane probes wrapping a
three-cycle. To represent these configurations, let us proceed as in section 3 and
take the following set of worldvolume coordinates
$\xi^{\mu}\,=\,(x^0,x^1,x^2,\theta_1,\phi_1,\psi)$, with 
$\theta_2=\theta_2(\theta_1,\phi_1)$, $\phi_2=\phi_2(\theta_1,\phi_1)$ and the
remaining two other coordinates $x^3$ and $r$ being constant. From the general
expression (\ref{Gammakappad5}), we obtain:
\beq
\Gamma_{\kappa}\,\epsilon\,=\,{i\over \sqrt{-g}}\,\,
{r^3\over L^3}\,\,\Gamma_{x^0x^1x^2}\,\gamma_{\theta_1\phi_1\psi}\,
\epsilon^*\,\,.
\eeq
The value of $\gamma_{\theta_1\phi_1\psi}\,\epsilon^*$ can be obtained by taking
the complex conjugate of eq. (\ref{gammathetaphipsi}), namely:
\beq
{18\over L^3}\,\,
\gamma_{\theta_1\phi_1\psi}\,\epsilon^*\,=\,-ic_1\Gamma_{\hat 3}\epsilon^*\,+\,
(c_2-ic_3)\,e^{i\psi}\,\Gamma_{1\hat 2\hat 3}\,\epsilon^*\,\,.
\label{gammathetaphipsi*}
\eeq
(The $c_i$ coefficients are given in eq. (\ref{d3cs})).
The only terms in (\ref{gammathetaphipsi*}) which could give rise to a projector
compatible with the $T^{1,1}$ conditions (\ref{tsprojections}) are the ones
containing the matrix $\Gamma_{1\hat 2\hat 3}$. Thus, we could  try to impose
the equation $c_1=0$. Notice, however, that the resulting kappa symmetry
projector is always going  to depend on the worldvolume coordinate $\psi$, due to
the $e^{i\psi}$ factor of the right-hand side of eq. (\ref{gammathetaphipsi*}). We
would have in this case a different projector for every point of the worldvolume,
which is, clearly, unacceptable. 

As an alternative solution to the problem just found, we could try to use a set of
worldvolume coordinates which does not include the coordinate $\psi$, \ie\ we will
consider $\psi$ as a constant scalar. After some calculations, one can convince
oneself that there is no consistent solution also in this approach. Therefore, we
are led to conclude that  these types of configurations are not supersymmetric.

As in section A.1, it is not difficult to find stable non-supersymmetric embeddings
of D5-branes wrapped on three-cycles. Indeed, for the election of worldvolume
coordinates and the ansatz for the scalar fields considered above, the lagrangian
density of the D5-brane probe is, up to irrelevant factors, the same as the one
obtained in section 3.1 for a D3-brane wrapping a three-cycle. We know from the
results of section 3.1 that the first-order equations (\ref{bpsd3}) imply the
fulfillment of the equations of motion. Thus, any solution of (\ref{bpsd3}) gives a
possible embedding of a D5-brane which wraps a three-cycle. Since, as in section
3.1.3, these embeddings saturate an energy bound, they are stable, despite of the
fact that they do not preserve any supersymmetry.

\subsection{The baryon vertex}

According to Witten's original argument \cite{Wittenbaryon}, the baryon vertex for
the Klebanov-Witten model must correspond to a D5-brane wrapped over the entire
$T^{1,1}$ space. Indeed, in this case the D5-brane captures the flux of the
Ramond-Ramond five-form $F^{(5)}$, which acts as a source for the worldvolume
electric field and, as a consequence, one must have fundamental strings emanating
from the D5-brane. The Dirac-Born-Infeld action must now necessarily include a
worldvolume gauge field $F$ and a Wess-Zumino term. It takes the form:
\beq
S\,=\,-T_5\,\int d^6\xi\,\sqrt{-\det (g+F)}\,-\,
T_5\int d^6\xi \,\,\,A\wedge F^{(5)}\,\,,
\label{baryonaction}
\eeq
where $A$ is the one-form potential for $F$, \ie\
$F=dA$ and $T_5$ is the tension of the D5-brane. For simplicity we are going to
take from now on the string coupling $g_s$ equal to one. Moreover, to obtain easily
the contribution of the Wess-Zumino term in (\ref{baryonaction}), it is useful to
rewrite the five-form $F^{(5)}$ of eq. (\ref{KW}) as:
\beq
F^{(5)}\,=\,{4L^4\over 108}\,\sin\theta_1\,\sin\theta_2\,
d\theta_1\wedge d\phi_1\wedge d\theta_2\wedge d\phi_2\wedge d\psi\,+\,
{\rm Hodge\,\,dual}\,\,. 
\label{F5}
\eeq
We shall take the following set of worldvolume coordinates 
\beq
\xi^{\mu}\,=\,(x^0,\theta_1,\phi_1,\theta_2,\phi_2,\psi)\,\,.
\eeq
For this election of the $\xi^{\mu}$'s, it is clear from the expression of 
$F^{(5)}$ displayed in eq. (\ref{F5}) that the one-form potential $A$ must have the
form $A=A_0dx^0$. Actually, we will adopt an ansatz in which the radial
coordinate $r$ and the gauge potential $A_0$ depend only on the angle $\theta_1$,
\ie:
\beq
r=r(\theta_1)\,\,,
\,\,\,\,\,\,\,\,\,\,\,\,
A_0\,=\,A_0(\theta_1)\,\,.
\label{baryonansatz}
\eeq
Notice that, within this ansatz, a 
worldvolume electric field $F_{x^0\theta_1}=-\partial_{\theta_1} A_0$
is switched on along the $\theta_1$ direction. 
The action  (\ref{baryonaction}) for such a configuration can be written as:
\beq
S\,=\,{T_5 L^4\over 108}\,\,V_4\,\,\int dx^0 d\theta_1\,\,
{\cal L}_{eff}\,\,,
\eeq
where $V_4$ is
\beq
V_4\,=\,\int d\phi_1 d\theta_2 d\phi_2 d\psi\sin\theta_2\,=\,32\pi^3\,\,,
\eeq
and the effective lagrangian density ${\cal L}_{eff}$  is given by:
\beq
{\cal L}_{eff}\,=\,-\sin\theta_1\,
\sqrt{\,{r^2\over 6}\,+\,(r')^2\,-(F_{x^0\theta_1})^2}\,-\,4\sin\theta_1 \,A_0\,\,.
\eeq
Let us now introduce the displacement field, defined as:
\beq
D(\theta_1)\equiv {\partial {\cal L}_{eff}\over \partial F_{x^0\theta_1}}\,=\,
\sin\theta_1\,
{F_{x^0\theta_1}\over \sqrt{\,{r^2\over 6}\,+\,(r')^2\,-(F_{x^0\theta_1})^2}}\,\,.
\label{displacement}
\eeq
The equation of motion for $A_0$ derived from ${\cal L}_{eff}$ gives the
Gauss' law for $D(\theta_1)$:
\beq
D'(\theta_1)\,=\,4\sin\theta_1\,\,,
\eeq
which can be immediately integrated, namely:
\beq
D(\theta_1)\,=\,-4\cos\theta_1+{\rm constant}\,\,.
\label{D}
\eeq
Integrating by parts the Wess-Zumino term in ${\cal L}_{eff}$ we can substitute  the
gauge potential $A_0$ by its canonically conjugate momentum $D$. This is of course
equivalent to performing a Legendre transformation.
The resulting  hamiltonian is:
\beq
H\,=\,{T_5 L^4\over 108}\,\,V_4\,\,\int  d\theta_1\,\,
{\cal H}\,\,,
\eeq
where ${\cal H}$ is given by:
\beq
{\cal H}\,=\,\sin\theta_1\,
\sqrt{\,{r^2\over 6}\,+\,(r')^2\,-(F_{x^0\theta_1})^2}\,+\,D(\theta_1)\,
F_{x^0\theta_1}\,\,.
\eeq
Since $F_{x^0\theta_1}$ and $D(\theta_1)$ are related by (\ref{displacement}), we
can eliminate in
${\cal H}$ the electric field $F_{x^0\theta_1}$  in favor of the displacement
$D(\theta_1)$, whose explicit expression is known from the integration of the Gauss'
law (see eq. (\ref{D})). Indeed, by inverting eq. (\ref{displacement}), we get:
\beq
F_{x^0\theta_1}\,=\,\sqrt{
{{r^2\over 6}\,+\,(r')^2\over D(\theta_1)^2\,+\,\sin^2\theta_1}}\,\,\,\,
D(\theta_1)\,\,,
\eeq
and, by using this relation, the hamiltonian density becomes:
\beq
{\cal H}\,=\,\sqrt{ D(\theta_1)^2\,+\,\sin^2\theta_1}\,\,\,\,
\sqrt{{r^2\over 6}\,+\,(r')^2}\,\,.
\eeq
The solutions of the equations of motion are the functions $r(\theta)$ which
minimize the above energy functional. Actually, the Euler-Lagrange equations derived
from
${\cal H}$ are rather involved and we will not try to find directly an analytical
solution. Instead, what we have tried is to apply the method used in ref.
\cite{severalbaryon} for the analysis of the baryon vertex in the   $AdS_5\times
S^5$  background. In ref. \cite{severalbaryon}, a first-order differential equation 
for $r(\theta)$ was found. The solutions of this equation also solve the
second-order Euler-Lagrange equation and saturate an energy bound. Despite the
similarity of our system with the one studied in ref. \cite{severalbaryon}, we have
not been able to find the first-order equation and the corresponding energy bound.
This fact suggests that the baryon vertex configurations in the 
$AdS_5\times T^{1,1}$ geometry are not supersymmetric. To confirm this result, let
us consider the kappa symmetry equation for this case. 
Using  the general expression  of $\Gamma_{\kappa}$ written in ref.
\cite{swedes} one obtains
\beq
\Gamma_{\kappa}\,\epsilon\,=\,-{i\over \sqrt{-\det (\,g+F\,)}}\,\,\Bigg[\,
{r\over L}\,\Gamma_{x^0}\,\gamma_{\theta_1\phi_1\theta_2\phi_2\psi}\,\,\epsilon^*\,-\,
F_{x^0\theta_1}\,\gamma_{\phi_1\theta_2\phi_2\psi}\,\,\epsilon\,\Bigg]\,\,.
\eeq
Moreover, since
\bear
&&\gamma_{\theta_1\phi_1\theta_2\phi_2\psi}\,\,\epsilon^*\,=\,
-{L^5\over 108}\,\sin\theta_1\sin\theta_2\,\Big(\,\Gamma_{\hat 3}\,+\,
\sqrt{6}\,\,{r'\over r}\,\Gamma_{r1\hat 3}\,\Big)\,\epsilon^*\,\,,\rc\rc
&&\gamma_{\phi_1\theta_2\phi_2\psi}\,\,\epsilon\,=\,
-{L^4\over 18\sqrt{6}}\,\sin\theta_1\sin\theta_2\,\Gamma_{1\hat 3}\,\epsilon\,\,,
\eear
we have
\bear
&&\Gamma_{\kappa}\,\epsilon\,=\,
{i\over \sqrt{-\det (\,g+F\,)}}\,L^4\,\sin\theta_1\sin\theta_2\,\Bigg[\,
{r\over 108}\,\Gamma_{x^0}\Gamma_{\hat 3}\,\epsilon^*\,+\,
{1\over 18\sqrt{6}}\,\Big(\,r'\,\Gamma_{x^0r1\hat 3}\,\epsilon^*\,-\,
F_{x^0\theta_1}\,\Gamma_{1\hat 3}\,\epsilon\,\Big)\Bigg]\,\,.\rc\rc
\label{baryongammakappa}
\eear
The kappa symmetry analysis of the baryon vertex in the $AdS_5\times S^5$ geometry
was performed in ref. \cite{Susybaryon}. The general strategy followed in 
\cite{Susybaryon} to solve the $\Gamma_{\kappa}\,\epsilon=\epsilon$ equation was
to try to impose an extra projection on  the spinor  in such a way that the
contributions of the worldvolume gauge field $F_{x^0\theta_1}$ and of $r'$ cancel
with each other. In our case it is clear that, by requiring that  
$\Gamma_{x^0r}\epsilon^*\,=\,-\epsilon$ the last two terms on the right-hand
side of eq. (\ref{baryongammakappa}) cancel with each other  if 
$F_{x^0\theta_1}=-r'$. Notice also that the condition
$\Gamma_{x^0r}\epsilon^*\,=\,-\epsilon$ corresponds to having fundamental strings along
the radial direction, which is just what we expect for a baryon vertex
configuration. Moreover, as can be checked by using the fact
that $\epsilon$ has fixed ten-dimensional chirality,  this extra projection is 
equivalent to require that $i\Gamma_{x^0\hat 3}\epsilon^*\,=\,\epsilon$, which in
turn is essential to satisfy the $\Gamma_{\kappa}\,\epsilon=\epsilon$ equation.
However, the fundamental string projection $\Gamma_{x^0r}\epsilon^*\,=\,-\epsilon$
is not consistent with the conditions (\ref{tsprojections}) satisfied by the Killing
spinors and, thus, it cannot be imposed to the $\epsilon$'s. Therefore, as
suspected, we are not able to solve the $\Gamma_{\kappa}\,\epsilon=\epsilon$
equation and we are led to conclude that the baryon vertex configuration breaks 
supersymmetry completely. Actually, from this incompatibility argument we see that
this conclusion is more general than the particular ansatz (\ref{baryonansatz})
that we have chosen.

\subsection{More spacetime filling D7-branes}

The election of worldvolume coordinates for the D7-brane probe that we have made in
section 5.1 might seem arbitrary. For this reason we will adopt here a different
point of view and take the following set of worldvolume coordinates
\beq
\xi^{\mu}\,=\,(x^0,x^1,x^2,x^3,\theta_1,\phi_1,\psi,r)\,\,,
\eeq
and we will consider configurations in which 
$\theta_2=\theta_2(\theta_1,\phi_1)$, $\phi_2=\phi_2(\theta_1,\phi_1)$. 
In this case, the kappa symmetry matrix $\Gamma_{\kappa}$  takes the form:
\beq
\Gamma_{\kappa}\,=\,-{i\over \sqrt{-g}}\,\,
{r^4\over L^4}\,\,\Gamma_{x^0x^1x^2x^3}\,\gamma_{\theta_1\phi_1\psi r}\,\,.
\eeq
Acting on a spinor $\epsilon_+$ such that $\Gamma_*\epsilon_+=\epsilon_+$ (with
$\Gamma_*$ defined in eq. (\ref{gamma*})), and using eq. (\ref{rhat3}) and eq.
(\ref{gammathetaphipsi}), one can demonstrate that:
\beq
{18r\over L^4}\,\,
\gamma_{\theta_1\phi_1\psi r}\,\epsilon_+\,=\,-c_1\epsilon_+\,+\,
i(c_2+ic_3)\,e^{-i\psi}\,\Gamma_{1\hat 2}\,\epsilon_+\,\,,
\eeq
where $c_1$, $c_2$ and $c_3$ are given in eq. (\ref{d3cs}). The compatibility of
the kappa symmetry projection  and eq. (\ref{tsprojections}) requires imposing
$c_2=c_3=0$, \ie\ the first-order differential equations (\ref{bpsd3}), whose
general solution was found in section 3.1.1 (eq. (\ref{generalholo})).
Since $\sqrt{-g}_{|BPS}={r^3\over 18}\,c_{1_{|BPS}}$, one concludes that 
$\Gamma_{{\kappa}_{|BPS}}\,\epsilon_+=\epsilon_+$, \ie\ the embeddings which satisfy
eq. (\ref{bpsd3}) preserve four supersymmetries. Moreover, it is also immediate
that the results of section 3.1.3 carry over to this case and, as a consequence,
one can establish a bound for the energy, which is saturated by the solutions of 
(\ref{bpsd3}). In
particular, the $m$-winding solutions (\ref{mwindingphi}) and (\ref{mwindingtheta})
can be represented by the polynomial equation (\ref{d3pol}) (as in section 5.1),
but now subjected to the conditions (\ref{d3polconditions}).

Notice that, in the BPS configurations just described, the D7-brane worldvolume
extends infinitely in $r$ and wraps a compact three-cycle of the $T^{1,1}$ space. In
particular, the D7-brane reaches the origin at $r=0$, contrary to what happens to
some of the embeddings studied in section 5.1.
This type of embeddings have been studied in refs. \cite{Ouyang,BLPV}.

\subsection{D7-branes wrapped on $T^{1,1}$}

A D7-brane can wrap the entire $T^{1,1}$ space and extend along two other spatial
directions of $AdS_5$. In order to find out if supersymmetry can be preserved in
this setup, let us choose the following set of worldvolume coordinates: 
$\xi^{\mu}\,=\,(x^0,x^1,r,\theta_1,\phi_1,\theta_2,\phi_2,\psi)$. The remaining
cartesian coordinates $x^2$ and $x^3$ are scalars which, in principle,  can depend
on the
$\xi^{\mu}$'s. First of all, let us assume that $x^2$ and $x^3$ are
constant and let us study whether or not this embedding is supersymmetric. The
general equation (\ref{gammakappa}) for this case becomes:
\beq
\Gamma_{\kappa}\,=\,-{i\over \sqrt{-g}}\,\,
\gamma_{x^0x^1r\theta_1\phi_1\theta_2\phi_2\psi}\,\,.
\eeq
For the configuration that we are considering, one can demonstrate that
$\Gamma_{\kappa}$ acts on the Killing spinors as:
\beq
\Gamma_{\kappa}\,\epsilon\,=\,i\Gamma_{x^0x^1r\hat 3}\,\epsilon\,\,.
\eeq
This equation is clearly solved for a spinor $\epsilon_+\,=\,r^{1/2}\,\eta_{+}$ 
(see eq. (\ref{chiraladsspinor})), where $\eta_{+}$ satisfies the extra projection
\beq
i\Gamma_{x^0x^1r\hat 3}\,\eta_{+}\,=\,\eta_{+}\,\,.
\eeq
Thus, this configuration preserves two supersymmetries. Moreover, by allowing $x^2$
and $x^3$ to depend on the worldvolume coordinates, one can convince oneself that 
$x^2$ and $x^3$ must be necessarily constant in order to preserve some
supersymmetry and, thus, the only supersymmetric embeddings are just the ones
studied above.

\vskip 1cm
\renewcommand{\theequation}{\rm{B}.\arabic{equation}}
\setcounter{equation}{0}
\medskip
\section{Supersymmetry of the Klebanov-Strassler solution}
\medskip

The Killing spinors of a supergravity background are obtained by requiring the
vanishing of the supersymmetry variations of the fermionic fields of the theory. 
For type IIB Sugra with constant dilaton these supersymmetry variations
are \cite{SUSYIIB}:
\bear
&&\delta\lambda\,=\,-{i\over 24}\,F_{\mu_1\mu_2\mu_3}\,
\Gamma^{\mu_1\mu_2\mu_3}\,\epsilon\,\,,\rc\rc
&&\delta\psi_{\mu}\,=\,D_{\mu}\,\epsilon\,+\,{i\over 1920}\,
F_{\mu_1\cdots\mu_5}^{(5)}\,\Gamma^{\mu_1\cdots\mu_5}\Gamma_{\mu}\epsilon\,+\rc\rc
&&\,\,\,\,\,\,\,\,\,\,\,\,\,\,\,\,+\,{1\over 96} F_{\mu_1\mu_2\mu_3}\,
\big(\,\Gamma_{\mu}^{\,\,\,\mu_1\mu_2\mu_3}\,-\,
9\delta_{\mu}^{\mu_1}\,\,\Gamma^{\mu_2\mu_3}\,\big)\,\epsilon^{*}\,\,,
\label{sugra}
\eear
where $\lambda$($\psi$) is the dilatino (gravitino) and $F$ is the following
complex combination of the Neveu-Schwarz-Neveu-Schwarz ($H$) and Ramond-Ramond (
$F^{(3)}$) three-forms:
\beq
F_{\mu_1\mu_2\mu_3}\,=\,g_s^{-{1\over 2}}\,H_{\mu_1\mu_2\mu_3}\,+\,i
g_s^{{1\over 2}}\,F_{\mu_1\mu_2\mu_3}^{(3)}\,\,.
\eeq
In what follows we will find the spinors $\epsilon$ which make that 
$\delta\lambda=\delta\psi_{\mu}=0$ for several backgrounds of type IIB
supergravity. First of all we will consider the space obtained by performing a
direct multiplication of a four-dimensional Minkowski space and a deformed
conifold. This geometry solves the equations of motion of type IIB supergravity
without forms and we will be able to find a simplified expression for its
Killing spinors. Then, we will add D3-branes, that warp the geometry and introduce
a non-vanishing value for the Ramond-Ramond five-form 
$F^{(5)}$,  and the result  will be reflected on the Killing spinors as an extra
projection to be satisfied by them and the multiplication of $\epsilon$ by a power
of the warp factor. The Klebanov-Strassler solution is obtained by adding
three-forms to this last background and requiring that the Killing spinors remain
unchanged \cite{GranaPol,Gubser}. We will explicitly verify that this requirement
allows to determine the metric and forms of this solution.

\subsection{Killing spinors of the deformed conifold}
\medskip
We will start by introducing the metric of the deformed conifold. With this
purpose, let us define the following set of one-forms
\bear
&&g^{1}\,=\,-{1\over \sqrt{2}}\,(\sigma^2-w^2)\,\,,
\,\,\,\,\,\,\,\,\,\,\,\,
g^{2}\,=\,{1\over \sqrt{2}}\,(\sigma^1-w^1)\,\,,
\,\,\,\,\,\,\,\,\,\,\,\,
g^{3}\,=\,-{1\over \sqrt{2}}\,(\sigma^2+w^2)\,\,,\rc\rc
&&g^{4}\,=\,{1\over \sqrt{2}}\,(\sigma^1+w^1)\,\,,
\,\,\,\,\,\,\,\,\,\,\,\,
g^{5}\,=\,\sigma^3+w^3\,\,,
\eear
where the $\sigma^i$ have been defined in eq. (\ref{sigmaoneforms}) and the $w^j$
are the one-forms displayed in eq. (\ref{woneforms}). The metric of the deformed
conifold can be written as \cite{Candelas}:
\bear
ds^2_{6}={1\over 2}\,\mu^{{4\over 3}}\,K(\tau)
&\Bigg[&{1\over 3 K(\tau)^3}\,\big(\,d\tau^2\,+\, (g^5)^2\,)\,+\,
\cosh^2\big({\tau\over 2}\big)\,
\big(\,(g^3)^2\,+\, (g^4)^2\,)\,+\,\rc\rc
&&+\,\sinh^2\big({\tau\over 2}\big)\,\big(\,(g^1)^2\,+\, (g^2)^2\,)\,
\,\Bigg]\,\,,
\label{deformedconifold}
\eear
where $\tau$ is a radial coordinate, $\mu$ is the deformation parameter of the
conifold and the function $K(\tau)$ is given by:
\beq
K(\tau)\equiv {\Big(\sinh 2\tau\,-\,2\tau\,\Big)^{1\over 3}\over
2^{{1\over 3}}\,\sinh\tau}\,\,.
\eeq

Let us now consider the following ten-dimensional metric:
\beq
ds^2_{10}\,=\,dx^2_{1,3}\,+\,ds^2_6\,=\,
-(dx^0)^2+\cdots+(dx^3)^2\,+\,ds^2_6\,\,,
\label{10dconifold}
\eeq
where $ds^2_6$ is the metric of (\ref{deformedconifold}).
This metric determines a Ricci flat geometry which solves the equations of motion
of supergravity without forms and preserves eight supersymmetries. As shown in ref. 
\cite{Twist}, this solution can be naturally obtained by uplifting  a domain
wall in gauged eight-dimensional supergravity. Actually, this eight-dimensional
origin provides us with the insight to find the frame basis in which the Killing
spinors become (almost) constant. To illustrate this fact, 
let us rewrite the conifold metric (\ref{deformedconifold}) 
as it is obtained in ref. \cite{Twist}, namely:
\bear
ds^2_{6}={1\over 2}\,\mu^{{4\over 3}}\,K(\tau)
&\Bigg[&{1\over 3 K(\tau)^3}\,\Big(\,d\tau^2\,+\, (w^3+\sigma^3)^2\,\Big)\,
+\,{\sinh^2\tau\over 2\cosh\tau}\,\Big(\,(\sigma^1)^2+(\sigma^2)^2\,\Big)\,+\rc\rc
&&+\,{\cosh\tau\over 2}\Big[\,\Big(\,w^1+{\sigma^1\over \cosh\tau}\,\Big)^2\,+\,
\Big(\,w^2+{\sigma^2\over \cosh\tau}\,\Big)^2\,\Big]\,\Bigg]\,\,.
\eear
We now choose the following frame one-forms for the ten-dimensional metric 
(\ref{10dconifold}):
\bear
&&e^{x^i}\,=\,dx^i\,\,,\,\,\,\,\,\,\,\, (i=0,1,2,3)\,\,,
\,\,\,\,\,\,\,\,\,\,\,\,\,\,\,\,
e^{\tau}\,=\,{\mu^{{2\over 3}}\over \sqrt{6}\,K(\tau)}\,\,d\tau\,\,,\rc\rc
&&e^{i}\,=\,{\mu^{{2\over 3}}\,\sqrt{K(\tau)}\over 2}\,\,
{\sinh\tau\over \sqrt{\cosh\tau}}\,\,\sigma^i\,\,,
\,\,\,\,\,\,\,\, (i=1,2)\,\,,\rc\rc
&&e^{\hat i}\,=\,{\mu^{{2\over 3}}\,\sqrt{K(\tau)}\over 2}\,\,
\sqrt{\cosh\tau}\,\,\Big(\,w^i+{\sigma^i\over \cosh\tau}\,\Big)\,\,,
\,\,\,\,\,\,\,\, (i=1,2)\,\,,\rc\rc
&&e^{\hat 3}\,=\,{\mu^{{2\over 3}}\,\over \sqrt{6}\,\, K(\tau)}\,\,
(w^3+\sigma^3)\,\,.
\label{KSframe}
\eear
By computing the different components of the spin connection in the basis 
(\ref{KSframe}), and substituting the results on the right-hand side of eq. 
(\ref{sugra}), one can check that the Killing spinors are independent of the
angular coordinates of the conifold and must satisfy the following projections:
\bear
&&\Gamma_{12}\,\epsilon\,=\,-\Gamma_{\hat 1\hat 2}\,\epsilon\,\,,\rc\rc
&&\Gamma_{\tau \hat 1\hat 2 \hat 3}\,\epsilon\,=\,-(\cos\alpha\,+\,\sin\alpha\,
\Gamma_{\hat 1 1}\,)\,\epsilon\,\,,
\label{projection}
\eear
with $\alpha$ being \cite{Twist}:
\beq
\cos\alpha\,=\,{\sinh\tau\over \cosh\tau}\,\,,
\,\,\,\,\,\,\,\,\,\,\,\,\,\,
\sin\alpha\,=\,-{1\over \cosh\tau}\,\,.
\label{rotationangle}
\eeq
Let us solve eq. (\ref{projection}) by  representing $\epsilon$ as:
\beq
\epsilon\,=\,e^{-{\alpha\over 2}\,\Gamma_{\hat 1 1}}\,\,\eta\,\,,
\label{deformedspinor}
\eeq
where $\eta$ is a  spinor satisfying the projections
\beq
\Gamma_{\tau \hat 1\hat 2 \hat 3}\,\eta\,=\,-\eta\,\,,
\,\,\,\,\,\,\,\,\,\,\,\,\,\,
\Gamma_{12}\,\eta\,=\,-\Gamma_{\hat 1\hat 2}\,\eta\,\,.
\label{etaproj}
\eeq
By substituting the representation (\ref{deformedspinor}) on the supergravity
variations (\ref{sugra}) one readily proves that $\eta$ is constant. Thus, the only
dependence of $\epsilon$ on the coordinates comes from the angle $\alpha$, which is
only a function of the radial coordinate $\tau$ and is such that 
$\alpha\to 0$ in the ultraviolet region $\tau\to \infty$ (see eq.
(\ref{rotationangle})). Therefore, the spinor $\epsilon$ at a finite value of
$\tau$ can be reconstructed from its asymptotic value $\eta$ by means of a
$\tau$-dependent rotation with angle $\alpha$. Notice that
the algebraic conditions  (\ref{etaproj}) determine eight spinors $\eta$, as
previously claimed.

\subsection{D3-branes on the deformed conifold }
\medskip
Let us now add D3-branes to the deformed conifold solution. This amounts 
to taking  a metric of the type
\beq
ds^2_{10}\,=\,h^{-{1\over 2}}\,dx^2_{1,3}\,+\,h^{{1\over 2}}\,ds^2_6\,\,,
\label{warpedmetric}
\eeq
where $h(\tau)$ is a warp factor, and  a Ramond-Ramond five-form of the form:
\beq
g_s\,F^{(5)}\,=\,d^4x\,\wedge dh^{-1}\,+\,{\rm Hodge\,\,\, dual}\,\,.
\label{KSfiveform}
\eeq
To determine the Killing spinors for this metric, we will use a frame
such as the one in (\ref{KSframe}), which now includes the corresponding powers of
the warp factor $h$.  By computing the contributions of the warp factor and
$F^{(5)}$ to the gravitino variation in (\ref{sugra}), one readily realizes that,
in addition to (\ref{projection}),  the Killing spinors $\epsilon$ for this
background satisfy an additional projection, namely:
\beq
\Gamma_{x^0x^1x^2x^3}\,\epsilon\,=\,-i\epsilon\,\,.
\label{d3projection}
\eeq
Moreover, since the type IIB spinors $\epsilon$ have fixed ten-dimensional chirality
\beq
\Gamma_{x^0x^1x^2x^3}\,\Gamma_{\tau 12 \hat 1\hat 2\hat 3}\,\epsilon\,=\,-
\epsilon\,\,,
\eeq
we have now
\beq
\Gamma_{\tau\hat 3}\,\epsilon\,=\,-i\epsilon\,\,,
\eeq
where we have used  (\ref{d3projection}) and the first equation in
(\ref{projection}). By combining this last equation with the relations in
(\ref{projection}) we obtain
\beq
\Gamma_{12}\,\epsilon\,=\,-\Gamma_{\hat 1\hat 2}\,\epsilon\,=\,i\,
(\cos\alpha\,+\,\sin\alpha \,
\Gamma_{\hat 1 1}\,)\,\epsilon\,\,.
\eeq
The Killing spinors in this case can be represented as:
\beq
\epsilon\,=\,h^{-{1\over 8}}\,e^{-{\alpha\over 2}\,\Gamma_{\hat 1 1}}\,\,\eta\,\,,
\label{KSspinors}
\eeq
where $\eta$ is a constant spinor that satisfies the projections (\ref{etaproj}),
together with
\beq
\Gamma_{x^0x^1x^2x^3}\,\eta\,=\,-i\eta\,\,.
\label{d3projectioneta}
\eeq
Notice that the algebraic conditions (\ref{etaproj}) and (\ref{d3projectioneta})
determine four spinors.

\subsection{Klebanov-Strassler solutions}
\medskip
Let us now add three-form gauge fields $H$ and $F^{(3)}$ to the solution of section
B.2. The metric and five-form will still be given by eqs. (\ref{warpedmetric}) and 
(\ref{KSfiveform}), while the three-forms will be parametrized by means of the
ansatz of ref. \cite{KS}, in which $F^{(3)}$ is written as:
\beq
F^{(3)}\,=\,{M\alpha'\over 2}\,\,\Big[\,
g^3\wedge g^4\wedge g^5\,+\,d\,\big(\,F(\tau)\,
(\,g^1\wedge g^3\,+\,g^2\wedge g^4\,)\,\big)\,\Big]\,\,.
\label{KSF3}
\eeq
In eq. (\ref{KSF3}) $M$ is a constant and $F^{(3)}$
is given in terms  of a single radial function $F(\tau)$. The  three-form $H$ is
obtained from a two-form potential $B$ as $H=dB$. The  ansatz of ref. \cite{KS} for
$B$ depends on two functions $f(\tau)$ and $k(\tau)$ as:
\beq
B\,=\,{g_s M\alpha'\over 2}\,\,\Big[\,
f(\tau)\,g^1\wedge g^2\,+\,k(\tau)\,g^3\wedge g^4\,\Big]\,\,.
\eeq
To determine the functions $F$, $f$ and $k$,
we shall impose to the Killing spinors the same projections as in the D3-brane case
and we will require that the contributions of the three-forms to the SUSY variation
of the dilatino and gravitino vanish by themselves. Thus, the addition of the
three-forms will not change the Killing spinors. From the condition
$\delta\lambda=0$ we get the following first-order equation
\beq
2F'\,+\,\coth\Big({\tau\over 2}\Big) \,f'\,-\,\tanh\Big({\tau\over 2}\Big) \,k'\,+\,
2\coth\tau\,F\,+\,f-k\,=\,\tanh\Big({\tau\over 2}\Big)\,\,.
\label{equno}
\eeq
Moreover, from $\delta\psi_{\hat 3}=0$ we get the differential equation
\beq
F'\,=\,{1\over 2}\,\coth\Big({\tau\over 2}\Big) \,f'\,-\,
{1\over 2}\,\tanh\Big({\tau\over 2}\Big) \,k'\,\,,
\label{eqcinco}
\eeq
and the following algebraic relation between the functions entering the  ansatz
\beq
2\coth\tau\,F\,+\,k\,-\,f\,=\,\tanh\Big({\tau\over 2}\Big)\,\,.
\label{eqseis}
\eeq
Plugging eqs. (\ref{eqcinco}) and (\ref{eqseis}) in (\ref{equno}), one immediately gets
\beq
F'\,=\,{k-f\over 2}\,\,,
\label{KSuno}
\eeq
which is one of the first-order equations found in ref. \cite{KS}. 
By inserting eq. (\ref{KSuno}) into eq. (\ref{eqseis}), we get the following
differential equation which only involves $F$ and its first derivative
\beq
F'\,+\,\coth\tau\,F\,=\,{1\over 2}\,\tanh\Big({\tau\over 2}\Big)\,\,.
\label{newdif}
\eeq
This equation can be integrated easily and the explicit form of $F$ can be obtained
(see below). 
By substituting the value of $F'$ given by eq. (\ref{newdif}) into eq. 
(\ref{eqcinco}), we get:
\beq
\coth\Big({\tau\over 2}\Big) \,f'\,-\,\tanh\Big({\tau\over 2}\Big) \,k'\,=\,
-2\coth\tau\,F\,+\,\tanh\Big({\tau\over 2}\Big)\,\,.
\label{eqcinconew}
\eeq
More equations are obtained from the SUSY variations of other components of the
gravitino. Indeed,  from the condition $\delta\psi_1=0$ we obtain two other
equations
\bear
&&f'-k'-2F+1-2\coth\tau\,\Big[\,F'\,+\,{k-f\over 2}\,\Big]\,=\,0\,\,,
\label{eqtres}\\ \rc
&&\coth^2\Big({\tau\over 2}\Big) \,f'\,-\,\tanh^2\Big({\tau\over 2}\Big) \,k'
\,-\,2(\,\coth^2\tau+{\rm csch}^2\tau\,)F\,+\,\rc\rc
&&\,\,\,\,\,\,\,\,\,\,\,\,\,+\,\tanh^2\Big({\tau\over 2}\Big)\,-\,
2\coth\tau\,\Big[\,F'\,+\,{k-f\over 2}\,\Big]\,=\,0\,\,.
\label{eqcuatro}
\eear
Notice that, by combining (\ref{KSuno})  and  (\ref{newdif}), we get:
\beq
2F'\,+\,{k-f\over 2}\,=\,2F'\,=\,\tanh\Big({\tau\over 2}\Big)\,-\,2\coth\tau\,F\,\,.
\eeq
Let us now use this result to compute the last term on the left-hand side of eq.
(\ref{eqtres}). After some calculation, eq. 
(\ref{eqtres}) becomes
\beq
f'-k'\,=\,-\Big[\,\coth^2\Big({\tau\over 2}\Big)\,+\, \tanh^2\Big({\tau\over 2}\Big)
\,\Big]\,F\,+\,\tanh^2\Big({\tau\over 2}\Big)\,\,.
\label{eqtresnew}
\eeq
By combining eqs. (\ref{eqcinconew})  and  (\ref{eqtresnew}) one can get $f'$ and $k'$
as functions of $F$. The result is
\beq
f'\,=\,(\,1-F\,)\,\tanh^2\Big({\tau\over 2}\Big)\,\,,
\,\,\,\,\,\,\,\,\,\,\,\,\,\,\,\,
k'\,=\,F\,\coth^2\Big({\tau\over 2}\Big)\,\,.
\label{KSdos}
\eeq
Eqs. (\ref{KSuno})  and  (\ref{KSdos}) constitute the first-order system of ref.
\cite{KS}. Apart from this system, we have obtained, in addition, a new differential
equation (eq. (\ref{newdif})) or, alternatively, the algebraic relation
(\ref{eqseis}). It can be checked that eq.  (\ref{eqcuatro}) follows from
(\ref{KSuno}), (\ref{KSdos}) and (\ref{newdif}). Moreover the SUSY variations of the
remaining components of the gravitino also cancel as a consequence of these
equations. 

Eq. (\ref{newdif}) can be easily integrated by the method of variation of constants.
The result is:
\beq
F\,=\,{1\over 2}\,\,{\sinh\tau\,-\,\tau\over \sinh\tau}\,+\,
{A\over \sinh\tau}\,\,,
\eeq
where $A$ is a constant. By requiring regularity of $F$ at $\tau=0$ we fix the constant
$A$ to the value $A=0$. It is then immediate to integrate the first-order equations
for $f$ and $k$. The result is the same as in ref. \cite{KS}, namely:
\bear
&&F\,=\,{1\over 2}\,\,{\sinh\tau\,-\,\tau\over \sinh\tau}\,\,,\rc\rc
&&f\,=\,{1\over 2}\,\,{\tau\coth\tau\,-\,1\over \sinh\tau}\,
(\,\cosh\tau-1\,)\,\,,\rc\rc
&&k\,=\,{1\over 2}\,\,{\tau\coth\tau\,-\,1\over \sinh\tau}\,
(\,\cosh\tau+1\,)\,\,.
\eear
Thus, the requirement of preserving the same supersymmetries as in the solution
corresponding to a D3-brane at the tip of a deformed conifold fixes the values of
the three-forms to those found in ref. \cite{KS}. Therefore, we conclude that the
Killing spinors of the Klebanov-Strassler background are given by eq.
(\ref{KSspinors}), with $\eta$ satisfying the projections (\ref{etaproj}) and
(\ref{d3projectioneta}).

\end{document}